\documentclass[12pt]{article}
\newcommand{\noop}[1]{}

\usepackage{comment}

\usepackage[usenames,dvipsnames,svgnames,table,x11names]{xcolor}
\usepackage[percent]{overpic}
 \usepackage[utf8]{inputenc} 
\usepackage{bbm,mathrsfs}
\usepackage{geometry}
\usepackage{anyfontsize}
\usepackage{t1enc}
\usepackage{hyperref}
\usepackage{braket}
\usepackage{packages-susy}
\usepackage{definitions}
\usepackage{simeontex-susy , macros-susy , nicefrac}
\usepackage{array} 
\usepackage{multirow} 
\usepackage{colortbl} 
\usepackage{arydshln} 
\usepackage{stfloats}  
\usepackage{rotating} 

\usepackage{pifont}

\def\rcy{\rowcolor{black!25!yellow!10}}
\def\rcr{\rowcolor{red!07}}
\def\cN{{\cal N}}

\usepackage{cite}

\usepackage{physics}
\usepackage{mathtools}
\hypersetup{
     colorlinks   = true,
     citecolor    = magenta
}

\def\outt#1{{}}

\renewcommand{\exp}[1]{{\rm exp}\{#1\}}

\renewcommand{\ll}{_}

\def\muchlessthan{< \hskip-.05in <}
\def\muchgreaterthan{> \hskip-.05in >}

\usepackage[affil-it]{authblk}

\renewcommand{\IR}{{\mathbb{R}}}

 \makeatletter
    
    \@addtoreset{equation}{section}
  \makeatother

\makeatletter
\g@addto@macro\bfseries{\boldmath}
\makeatother

\def\rdots{\redd{\circ\circ\circ}}

\def\D{\Delta}

\newenvironment{PurpleEnv}%
{\color{Purple}}%
{\color{Black}}
{\color{Blue}}%
{\color{Black}}
{\color{Red}}%
{\color{Black}}

\def\purple#1{\begin{PurpleEnv} {#1} \end{PurpleEnv}}

\def\tb{{\bar{\tau}}}

\def\cob{{\overline{\co}}}

\def\rwa#1{\reduwave{#1}}

\def\ampname{{\cal Y}}
\def\GGO{& ~ & ~ }
\def\GGO{}
\usepackage{subcaption}

\def\JJM{{\textcolor{BlueViolet}{{{\cal J}}}}}
\begin{document}
\hypersetup{linkcolor=black}

\date{\today}

\begin{titlepage}
\begin{flushright}
IPMU17-0143\\
CALT-TH-2017-059\\
\end{flushright}
\vspace{8 mm}
\begin{center}
   {\large \bf On the Large $R$-charge Expansion} \\ \vspace{2 mm}{\bf \large in} \\  \vspace{2 mm}
  {\bf \large ${\cal N}=2$ Superconformal Field Theories }
\end{center}
\vspace{2 mm}
\begin{center}
{Simeon Hellerman and Shunsuke Maeda} \\
\vspace{6mm}
{\it Kavli Institute for the Physics and Mathematics of the Universe (WPI)\\
The University of Tokyo \\
 Kashiwa, Chiba  277-8582, Japan\\}
%
\vspace{6mm}

\end{center}
\vspace{-4 mm}
\begin{center}
{\large Abstract}
\end{center}
\noindent


\noindent
In this note we study
two point functions of Coulomb branch
chiral ring elements with large $R$-charge, in quantum
field theories with ${\cal N} = 2$ superconformal symmetry in four spacetime dimensions.
Focusing on the case of
one-dimensional Coulomb branch,
we use the effective-field-theoretic methods of
\cite{Hellerman:2017veg},
to
estimate the two-point correlation function $\ampname\ll n \equiv  |x - y|\uu{ 2 n\D\ll\co}\cc\left< (\co (x) )\uu n
\cc  (\bar{\co}(y) )\uu n \right>$
%
 in the limit where the operator insertion $\co\uu n$ has large total $R$-charge $\JJM = n\D\ll\co$.  We show
 that $\ampname\ll n$ has
 a nontrivial but universal asymptotic expansion at large $\JJM$, of
the form
\begin{align*}
\begin{split}
\ampname\ll n = 
\JJM ! \cc \left({{|{\bf N}\ll\co|}\over{2\pi}} \right)\uu {2 \JJM} \cc  \JJM\uu\a \cc  \tilde{\ampname}\ll n\ ,
\end{split}
\end{align*}
where  $\tilde{\ampname}\ll n$ approaches a constant as $n\to\infty$, and ${\bf N}\ll\co$ is an $n$-independent constant describing
on the normalization of the operator relative to the effective Abelian
gauge coupling.
The exponent $\a$ is a positive
number proportional to the difference between the $a$-anomaly coefficient of the underlying CFT and that
of the effective theory of the Coulomb branch.  For Lagrangian SCFT,
we check our predictions for the logarithm ${\cal B}\ll n = \log (\ampname\ll n)$, up to and including order $\log(\JJM)$ against exact
results from supersymmetric
localization\cite{Baggio:2014ioa,Baggio:2014sna,Baggio:2015vxa,Gerchkovitz2017}.  In the case of ${\cal N} = 4$ we find precise agreement and in the case ${\cal N} = 2$ we find  reasonably good numerical agreement at
$\JJM \simeq 60$ using the no-instanton approximation to the $S^4$ partition function.
 We also give predictions for the growth of two-point functions in all rank-one SCFT in the classification of \cite{Argyres:2015ffa,Argyres:2015gha, Argyres:2016xmc,Argyres:2016xua}.  In this way, we show the large-$R$-charge expansion serves as a bridge from the world of unbroken superconformal symmetry, OPE data, and bootstraps, to the
world of the low-energy
dynamics of the moduli space of vacua.


\vspace{.5cm}
\begin{flushleft}
\today
\end{flushleft}
\end{titlepage}

\tableofcontents\hypersetup{linkcolor=SeaGreen4}
\newpage
\numberwithin{equation}{section}

\section{Introduction}

Recently  there has been development of the properties of conformal field theories (CFTs) with global charges,
in states of large quantum number 
\cite{Hellerman:2015nra,Alvarez-Gaume:2016vff,Monin:2016jmo,Loukas:2016ckj,Hellerman:2017efx,Hellerman:2017veg,Loukas:2017lof,Banerjee:2017fcx,Fitzpatrick:2012yx,Komargodski:2012ek,Li:2015itl,Li:2015rfa,Kaviraj:2015cxa,Kaviraj:2015xsa,Dey:2017fab,Alday:2015ewa,Alday:2016njk,Alday:2016jfr,Hellerman:2016hnf, Hellerman:2014cba,Hellerman:2013kba,Costa:2012cb,Caron-Huot:2016icg,Sever:2017ylk}.\footnote{Another paper involving the subject of conserved charges includes \cite{Son:2005rv}.
We thank V. Rychkov for pointing out this paper to us.}
Many large quantum number analyses have studied the asymptotic
expansion of operator dimensions at large quantum number $J$ (either an internal symmetry such as spin, or an internal
global symmetry such as a $U(n)$ or $O(n)$ symmetry), in negative powers of the total charge $J$.  In many cases these relations have been checked and remarkable agreement has been found among various methods to study such regimes: conformal bootstrap\footnote{For a review of the conformal bootstrap, including the dramatic progress \cite{Rattazzi:2008pe,El-Showk:2014dwa,ElShowk:2012ht} of the modern era, see \cite{Rychkov:2016iqz,Simmons-Duffin:2016gjk} and references therein.}, S-matrix bootstrap, Monte Carlo simulation, and the use of effective field theory.  Also, in \cite{Monin:2016jmo} three-point functions were studied and
a similar expansion was derived for OPE coefficients (equivalently, three-point functions) where at least one of the operators has large global charge $J$ (and therefore, automatically, at least two of the operators).

 The general pattern emerging from these works is that
the behavior of operator dimensions and OPE coefficients becomes simple in the limit of some large quantum number $J$
even in a strongly coupled system.   This pattern is interesting because its robustness appears to transcend the explanations for it in the analysis of any individual
case.  In some cases, such as large spin in a single plane in a CFT, the explanation lies in the conformal bootstrap and
looks particularly intuitive in an AdS holographic dual.  For systems with large global charge carried by bosonic
fields,
the most readily apparent explanations appear to involve a large-charge effective field theory, in which
the large-$J$ operator is well-approximated by a smooth classical solution in radial quantization.  These
large-quantum-number limits themselves appear to be special cases of an even more general situation, a "macroscopic limit" in which
one takes pure or mixed states or density matrices into some extreme direction in Hilbert space.  The "eigenstate thermalization
hypothesis" \cite{PhysRevE.50.888}, when it holds,
is perhaps the most famous example of this behavior.\footnote{We thank D. Jafferis and A. Zhiboedov for bringing this analogy to our attention and for communicating preliminary
results on a bootstrap derivation of some properties of the large-$J$
behavior in \cite{Hellerman:2015nra, Alvarez-Gaume:2016vff, Monin:2016jmo}
making use of Regge theory and the conceptual connection with the ETH \cite{Jafferis:2017zna, sashatalk}.}

In \cite{Hellerman:2017veg} the authors analyzed the large-$R$-charge expansion
for operator dimensions in a superconformal field theory with a one-complex-dimensional moduli space, and were able to quantize the effective theory on moduli space in radial quantization, in order to compute operator dimensions of BPS and near-BPS primary operators of large $R$-charge $\JJM$.  It was shown that the basic predictions of superconformal invariance (such as nonrenormalization of
the energies of BPS scalar primaries) can be recovered, and nontrivial predictions about semi-short states can be made
and verified via the superconformal index.  Additionally, one easily derives many nontrivial relations on the asymptotic expansion
on the energies of the low-lying non-BPS states, including nontrivial information about the $\JJM$-scaling (order $\JJM\uu{-3}$)
and sign (negative) of the leading correction to the first non-BPS primary dimension.

It is therefore natural to attempt to make further contact between the large-$J$ expansion and other methods that make maximal
use of superconformal symmetry.  To do this, one would like to find a set of observables associated with states of large $R$-charge which both has a nontrivial $\JJM^{-1}$ expansion, like the non-BPS energies in \cite{Hellerman:2017veg}, and is also controlled directly by exact superconformal symmetry.  The obvious candidate is the three-point functions of two BPS and one anti-BPS chiral primary scalar operators.  These three-point functions are equivalent to OPE coefficients of the chiral ring, and are therefore independent of $D$-terms
in the effective action.  At the same time, they have a nontrivial dependence on $\JJM$, unlike the operator dimensions of BPS states.
Due to their independence of $D$-terms, these three-point functions can in principle be computed exactly by supersymmetric
localization on $\Sfour$ and in some cases have been worked out explicitly \cite{Baggio:2014ioa,Baggio:2014sna,Baggio:2015vxa,Gerchkovitz2017,Rodriguez-Gomez:2016ijh,Dedushenko:2016jxl,Baggio:2016skg,Pini:2017ouj,Baggio:2017aww,Billo:2017glv,Agmon:2017lga} (See also an  earlier work \cite{Papadodimas:2009eu}). 

In this note we will compute three-point functions of chiral ring elements in theories with ${\cal N} \geq 2$ superconformal
symmetry in four spacetime dimensions, with a one-complex-dimensional Coulomb branch.  Such three-point
functions are more conveniently expressed as two-point functions 
\begin{align}
\begin{split}
\ampname\ll n\equiv \left< \co\uu n \cob
\vphantom{\co}
\uu n \right>
\ ,
\label{TwoPtAmpDef}
\end{split}
\end{align}
where for any operator $\co$ we abbreviate
\begin{align}
\begin{split}
\langle \co \cob \rangle \equiv  |x|\uu{+ 2 \D\ll\co}\cc \langle \co(x) \cob(0)\rangle\ll{\IR\uu 4}  ,
\label{RescalingAbbrevDef}
\end{split}
\end{align}
which is independent of $x$ in a CFT.  For a one-complex-dimensional Coulomb branch, its chiral ring is generated by a single element $\co$, which we take to be of dimension 
\bbb
\Delta\ll\co = |J\ll\co|\ .
\een{BPSFormulaForJOfCalO}
In equation \rr{BPSFormulaForJOfCalO} we\footnote{slightly nonstandardly but very conveniently} normalize the $R$-charge $J$ so that the ${\cal N} = 2$ supercharges $Q\ll\a\uu i$
have $J = - {1\over 2}$, the BPS bound for scalar operators is $\D\geq |J|$, and a free field $\phi$ has $\D\ll\phi = |J\ll\phi | = 1$.

The main focus of this paper is to show the large-$n$ behavior of the function $\ampname\ll n$ is universal, behaving
asymptotically as
\begin{align}
\begin{gathered}
\ampname\ll n = \eexp\left({{\cal B}\ll n}\right)\ ,    
\\
{\cal B}\ll n = \log\left [ \cc  \JJM ! \cc \right ] + b\ll{-1}\cc \JJM + \a\cc \log(\JJM)
   + O(n\uu 0)\ ,
\end{gathered}
\end{align}
where $\JJM$ is the total $R$-charge $\JJM = n\cc |J\ll\co| = n\D\ll\co$ of the operator insertion $\co\uu n$, 
and the remainder is bounded as $\JJM\to\infty$.

 The coefficient $b\ll {-1}$ is $n$-independent but depends on the normalization
of the metric on moduli space relative to
the normalization of the operator $\co$ itself.
The constant $\a$ is related to the
the $a$-coefficient in the Weyl anomaly. 
The definition of the normalization of the anomaly coefficient in turn depends on a convention, but the exponent
$\a$ has an absolute meaning, so we should express the value of $\a$ in a convention-independent way.  The value of $\a$ can be expressed as
\begin{align}
\begin{split}
\a \equiv  {{5 \Delta a}\over{12\cc a\lrm{favm}}}\cc\ ,
\end{split}\label{AlphaCoefficientValuePREVIEW}
\end{align}
where $\Delta a$ is the difference between the $a$-coefficient of the underlying CFT $a\lrm{CFT}$ and
the $a$-coefficient of the effective theory of massless
fields on moduli space, $a\lrm{EFT}$.  The
normalization $a\lrm{favm}$ in the denominator
denotes the $a$-anomaly of a free Abelian vector multiplet
of ${\cal N} = 2$ supersymmetry.  We have expressed
the value of $\a$ in this form in order to describe it
independent of normalization convention.  In one commonly
used convention of \cite{Anselmi:1997ys} by Anselmi, Erlich, Freedman and Johansen (AEFJ), the value of $a_{\mathrm{favm}}$ is $\nicefrac{5}{24}$, and so
\begin{align}
\begin{split}\label{alphaAEFJ}
\a = 2 (a\lrm{CFT} - a\lrm{EFT})\ups{\rm AEFJ}
\end{split}
\end{align}
in that convention.  In table \ref{tabletable}, we give a list of values for the $\a$ coefficient in all known ${\cal N} = 2$ superconformal
field theories whose moduli space at a generic point has only one Abelian vector multiplet.  For example,
${\cal N} = 4$ super-Yang--Mills theory with gauge algebra $\mathfrak{su}(2)$, has an $\a$-coefficient $\a = +1$.

As we have seen in \cite{Hellerman:2017veg}, the insertion of an operator of large $R$-charge creates
a state of large $R$-charge on $S\uu 3$ via radial quantization.  Though we will not be using
radial quantization or the $S\uu 3 \times ({\rm time})$ conformal frame at all
in the present paper, the underlying physics is the same, as is the reason for 
recovering a semiclassical description: The large-$n$ limit of two-point functions $\ampname\ll n$
is a large-$\JJM$ limit, in which the effective theory becomes weakly coupled on the infrared scale.

The leading term is contributed by the action of the classical solution created by the insertions $(\co(x))\uu n (\cob(y))\uu n$, the
term $b\ll{-1} \JJM$ depends on the normalization of the operator $\co$, and
the subleading term $\a\cc\log (\JJM)$, is contributed by the Wess--Zumino lagrangian in the effective theory on the Coulomb branch.  The remaining regular terms come from quantum corrections within the effective theory, as well as explicit higher-derivative terms in the effective action of the moduli space, with
unknown coefficients.  (Some low-derivative terms in effective actions have
been constructed \cite{Argyres:2003tg, Argyres:2004kg, Argyres:2004yp, Argyres:2008rna} but still there is no full classification even at low orders in the derivative expansion, let alone anything known about the coefficients of
higher-derivative terms even for simple ${\cal N} = 2$ SCFT.)
Quantum
corrections within the effective action itself start only
at order $\JJM\uu 0$; even those are summed up entirely by the free-field action.  That is, the only
quantum effects contributing at order $\JJM\uu 0$ are determinants in the free effective theory, and
these are completely summed up by Wick contractions of the free abelian vector multiplet scalar
describing the Coulomb branch.

\section{Large-$R$-charge expansion of two-point functions}
\label{LargeChargeExp}

Much of the setup of the calculation is similar to that
of \cite{Hellerman:2017veg}, and we shall refer the reader to consult
that paper to the extent the two calculations are more or less parallel.  Two differences include the dimensionality of spacetime ($D=4$ in the present paper versus $D=3$ in
\cite{Hellerman:2017veg}) and the amount of supersymmetry (eight Poincar\'e supercharges here versus only four in \cite{Hellerman:2017veg}), but these distinctions make little difference to the structure of the large-$J$ expansion, and we will mostly just
refer to the superspace analysis of \cite{Hellerman:2017veg}, drawing
attention to differences as they become relevant.

In particular, the case of
eight supercharges in $D=4$ can be seen as a special case of ${\cal N} = 1$ SUSY in $D=4$, which dimensionally reduces to ${\cal N} = 2$ SUSY in $D=3$, as in the case of the model
studied in \cite{Hellerman:2017veg}.

Our analysis will apply to all superconformal
theories in four dimensions with ${\cal N} \geq 2$ SUSY and a one-dimensional Coulomb branch.  We will only make use of the ${\cal N} = 2$ subalgebra, with ${\cal N} = 3,4$ theories being subsumed as special cases.  When we refer to dimensions of moduli spaces, we will always be using ${\cal N} = 2$ terminology, in terms of which {\it e.g.},  the $G = SU(N)$ super-Yang--Mills
theory with ${\cal N} = 4$ SUSY, can be thought of as an ${\cal N} = 2$ theory with gauge group $G$ and a single adjoint
hypermultiplet, so that the moduli space is described by $N-1$ vector multiplets and $N-1$ massless neutral hypermultiplets.

\subsection{Basics}
\heading{Two-point functions as three-point functions}

For ${\cal N} = 2$, $D = 4$ superconformal theories with one-dimensional Coulomb branch coordinatized by the chiral primary operator $\co$, 
the nonvanishing three-point functions are 
\begin{align}
\begin{split}
{\bf C}^{n\ll 1 ,~ n\ll 2,~ \overline{n\ll 1 + n\ll 2}} \equiv \left< \co\uu {n\ll 1} \co\uu{n\ll 2} \cc\cob\vphantom{\co}\uu{n\ll 1 + n\ll 2} \right>\ ,
\end{split}
\end{align}
where 
 we have suppressed the position dependence on the positions of the amplitude, as it is determined uniquely by the conformal Ward identity and the dimensions $n\ll i\cc
\D\ll \co \equiv  n\ll i \cc  |J\ll\co|$
 of the operators.  Explicitly,
we have used the abbreviation
\begin{align}
\begin{split}
\left< \co\uu {n\ll 1} \co\uu{n\ll 2} \cc\cob\vphantom{\co}\uu{n\ll 1 + n\ll 2} \right>
\equiv \left|x\ll 1 - y\right|\uu{+ 2 n\ll 1 \D\ll\co} \cc\left |x\ll 2 - y\right|\uu{+ 2 n\ll 2 \D\ll\co} \cc 
\left< \co\uu {n\ll 1}(x\ll 1) \cc  \co\uu{n\ll 2}(x\ll 2) \cc\cob\vphantom{\co}\uu{n\ll 1 + n\ll 2} (y)\right>
\end{split}
\end{align}
Chiral primaries (with the same sign of the $R$-charge)
have the special property that their OPE is nonsingular, and therefore their three-point functions can be reduced immediately to two-point functions, by taking the two like-charge chiral primaries in the correlator, to lie at the
same point, and we have
\begin{align}
\begin{split}\label{CalANDef}
 \ampname\ll n \equiv {\bf C}\uu{n\pr,~ n - n\pr ,~ \bar{n}} =
\left<\co\uu n \cob\vphantom{\co}\uu n \right>
\equiv
\left|x\right|^{2n\Delta_{\co}}
\left<
\co^n(x)\cob\vphantom{\co}^n(0)
\right>
 \ ,
\end{split}
\end{align}
independent of $n\pr$.

  At first sight, the notion of a meaningful normalization for
  a conformal two-point function seems unfamiliar, because one generally thinks of two-point functions as simply being equal to $1$.  However this normalization convention, while widely used, is not the natural one for elements of the chiral ring.  Once a set of generators for the chiral ring
  has been chosen, the higher operators generated from them algebraically, are defined principle by associativity, including
  their normalization.  That is,
  if one fixes the normalizations of chiral ring elements
  $\chi\ll 1$ and $\chi\ll 2$, one no longer has the
  freedom to unit-normalize the product $\chi\ll 3 \equiv \chi\ll 1 \chi\ll 2$.
  
  In the case of a one-dimensional chiral ring, the normalization of the generator $\co$ is arbitrary, but once it has
  been chosen, one no longer has the freedom to rescale the operators $\co\uu n$, and their two-point functions $\ampname\ll n$ have
  nontrivial dependence on  $n$ which is an output of the dynamics of the theory, related via \rr{CalANDef}
  to three-point functions.

\subsection{Free-field approximation}

\heading{Two-point functions on $\IR\uu 4$}

In the introduction we mentioned some differences from
the case of \cite{Hellerman:2017veg},
such as the amount of SUSY and the number of spacetime dimensions, that do not much alter the structure of the
calculation.  A more relevant difference, is that we will perform
our computation directly as a two-point correlator
on flat space $\IR\uu 4$, rather than in radial
quantization as we did in \cite{Hellerman:2017veg}.  This is because
the observable we are studying, the norm of the state
itself, is harder to see directly in radial quantization, as
the Hilbert space formalism normally begins by
taking the norm of the state as an input.  Thus
radial quantization is more directly useful for computing
the power law in the two-point function -- \it i.e., \rm the energy of the state -- than for computing the overall
normalization of the two-point function.  We could
of course compute these same observables 
in radial quantization as three-point functions, as
was done for the $O(2)$ model or other CFT
described at large charge by a conformal goldstone EFT, as described in generality in \cite{Monin:2016jmo}.
Checking that these two methods produce the same
result for $\ampname\ll n$ would be a valuable check on the
consistency of our approach, but we will not pursue it in
the present article.

\heading{Free effective field}

By assumption, our effective action contains a single vector multiplet, plus possibly massless hypermultiplets.  We shall
ignore the latter for the moment, as they will not affect the classical solution that controls the leading 
terms in the two-point function.  In discussing the vectormultiplet effective action,
we will mostly\footnote{With one particular exception: For the microscopic holomorphic gauge coupling in conformal ${\cal N} = 2$ SQCD, 
the references \cite{Seiberg:1994rs, Seiberg:1994aj}
define $\t = {{8\pi i}\over{g\sqd\lrm{YM}}} + {{\th}\over{\pi}}$ for reasons to do with duality and Dirac
quantization condition.  We will however use
the convention $\t = {{4\pi i}\over{g\sqd}} + {{\th}\over{2\pi}}$ for all gauge couplings, both microscopic and effective,
uniformly in the representation of the hypermultiplets.
This convention is more
commonly used recently, particularly in the literature on localization, \it e.g. \rm \cite{Pestun:2007rz} and works
making use of it.} follow the conventions of \cite{Seiberg:1994rs, Seiberg:1994aj}, and give explicit translation
of other quantities into the normalizations of \cite{Seiberg:1994rs, Seiberg:1994aj}.  The complexified gauge
coupling $\t$ is
\begin{align}
\begin{split}
\t\lrm{eff} \equiv {{4\pi i}\over{g\sqd\lrm{eff}}} + {{\th\lrm{eff}}\over{2\pi}} .
\end{split}
\end{align}
The degrees of freedom in a vector multiplet are an abelian gauge field, and neutral fermions and a neutral complex
scalar $A$.  In terms of the field $A$, the coupling $\t\lrm{eff}$ is given by 
\begin{align}\label{EffTauFromEffPrepot}
\begin{split}
\t\lrm{eff} = {\cal F}\ll{\rm eff}\prpr(A)\ ,
\end{split}
\end{align}
where ${\cal F}$ is the effective holomorphic prepotential
for the Abelian vector multiplet based on $A$.

The kinetic term for $A$ contains nontrivial dynamical information and is related to the
abelian gauge coupling, but we do not know anything a priori about the kinetic term for $A$, other than that
it respects the symmetries of the system.  The combination of $R$-symmetry and scale-invariance force
metric on moduli space to be flat.  Then the kinetic term for $A$ has to be
\begin{align}
\begin{gathered}\label{KineticTermForEffVMScalarA}
S\lrm{free}  \equiv \int \cc d\uu 4 x \cc
{\cal L}\lrm{free}\ ,
\\{\cal L}\lrm{free} = {{\Im(\t\lrm{eff})}\over{4\pi}}\cc (\pp\ll\m A)(\pp\ll\m \bar{A}) = g\uu{-2}\lrm{eff} \cc  (\pp\ll\m A)(\pp\ll\m \bar{A}).
\end{gathered}
\end{align} 
The complexified effective gauge coupling is related to the effective prepotential  by \rr{EffTauFromEffPrepot}
and must be constant as a function of $A$ in a conformal theory, so the effective prepotential is
\begin{align}
\begin{split}
{\cal F}(A) = {{\t\lrm{eff}\cc A\sqd}\over 2}\ .
\end{split}
\end{align}
The parameter $\t\lrm{eff}$ can depend on any marginal coupling parameters that may be present, but
not on the dynamical field $A$.

We can define a field with unit kinetic term by
\begin{align}
\begin{split}
\phi\lrm{unit} = g\uu{-1}\lrm{eff} \cc A = \sqrt{{\Im(\t\lrm{eff})}\over{4\pi}} \cc A
\end{split}\label{VMScalarAToUnitField|Phi}
\end{align}
so that the kinetic term is
\begin{align}
\begin{split}\label{UnitScalarKineticTermNormalizationDefPRECAP}
{\cal L}\lrm{free} \equiv\left |\pp\phi\lrm{unit}\right|\sqd = (\pp\ll\m\phb\lrm{unit})(\pp\ll\m\phi\lrm{unit})\ .
\end{split}
\end{align}
Note that the transformation  \rr{VMScalarAToUnitField|Phi} is holomorphic in $A$ and $\phi$ but not
in the background couplings controlling $\t\lrm{eff}$.
We will drop the subscript ${}\lrm{unit}$ for the remainder of this article except when potentially unclear.

\heading{Normalization of the effective scalar}

In order to evaluate the $n\uu {+1}$ term in ${\cal B}\ll n \equiv \log(\ampname\ll n)$, one would need
to relate the vector multiplet scalar $A(x)$,
to the generator 
$\co(x)$
of the Coulomb branch chiral ring,
of whose $n\uu{\underline{\rm th}}$ powers we are taking the two-point function.
We must have
\begin{align}
\begin{split}
\co \equiv ({\bf M}\ll\co\cc A)\uu{\D\ll\co}\ ,
\end{split}\label{CalOToPhiUnit}
\end{align}
for some constant ${\bf M}\ll\co$, where $\Delta\ll\co$
is the conformal dimension of $\co$.  The change of variables is locally holomorphic as 
a function of $A$, and the normalization constant ${\bf M}\ll\co$ should be holomorphic in all background fields as well, such a any
marginal directions in the space of couplings:
\begin{align}
\begin{split}
{\bf M}\ll\co = {\bf M}\ll\co(\t)\ , \qquad {{\partial {\bf M(\t)}}\over{\partial \tb}} = 0\ .
\end{split}
\end{align}
Defining ${\bf N}\ll\co \equiv {\bf N}_\co(\t,\tb)$ such that
\bbb
{\bf N}\ll\co \cc \phi ={\bf M}\ll\co \cc A\ ,
\eee
the quantities ${\bf N}\ll\co$ and ${\bf M}\ll\co$
are related by
\begin{align}
\begin{split}
{\bf N}\ll\co = g\lrm{eff}\cc {\bf M}\ll\co = \sqrt{{4\pi}\over{\Im(\t)}} \cc {\bf M}\ll\co\ .
\end{split}
\end{align}

Since we are assuming $\phi$ has unit kinetic term in the Lagrangian,
we cannot absorb ${\bf N}\ll\co$ into the definition of $\phi$.  
We could of course absorb ${\bf N}\ll\co$ into the definition of $\co$, but we might want to normalize $\co$
in some other way.  For instance, we might want to take $\co$ itself to have unit two-point function $\ampname\ll 2$.
For general ${\cal N} = 2$ theories with one-dimensional Coulomb branch, we do not know a simple way
to calculate ${\bf N}\ll\co$ given some preferred normalization of $\co$: This is an interesting problem
for future investigation of the large-$R$-charge limit.  
For particular theories, namely those with a marginal
coupling $\t\equiv \t\lrm{YM}$, we shall
be able to say more, and we will come to this
situation in a later section.

For now, however, we leave ${\bf M}\ll\co$ unspecified.
In terms of this factor, the map between $\co$ 
and $\phi$ can be written
\begin{align}
\begin{split}
\co = \left(  {\bf N}\ll\co  \right) \uu{\JJM}\cc
\phi\uu{\JJM}= 
\left( g\lrm{eff}\cc
{\bf M}\ll\co \right)\uu{\JJM}\cc
\phi\uu{\JJM} = \left(\sqrt{{4\pi}\over{\Im(\t)}}
\cc {\bf M}\ll\co \right) \uu{\JJM}\cc
\phi\uu{\JJM}\ .
\end{split}
\end{align}

\heading{Multivaluedness of the map between
$\co$ and $\phi$}

Note that the map between $\co$ and $A$ or $\phi$ is not one to one in general.  If $\D$ is an integer,
then the map from $\phi$ to $\co$ is single valued,
but it is only one-to-one if $\D\ll\co = 1$ which holds
only in a free theory.  If $\D\ll\co$ is not
an integer, the map from $\phi$ to $\co$ is
not even single-valued.\footnote{Though $\D\ll\co$ is
integer in Lagrangian theories, 
 there
are various non-Lagrangian rank-one theories (so-called  Argyres--Douglas theories \cite{Argyres:1995jj,Argyres:1995xn,Xie:2012hs})
with fractional $\Delta\ll\co$.  See, \it e.g. \rm table 1 of \cite{Argyres:2016xmc}.}

 The coordinate $\phi$ or $A$
should be thought of only as a local holomorphic coordinate here.   
 As long as we are away from the origin where
the effective theory is valid, the singularity should not
affect the validity of the effective action. 
 
 Treating $a$ or $\phi\lrm{unit}$ as a local coordinate
 should not affect the validity of our use of the
effective field theory, as it does not in the usual
manipulations of Seiberg-Witten theory.  In the case
where $\D\ll\co$ is noninteger, we may for simplicity
restrict ourselves
to the case,
where $\JJM$ is integer, so that our Wick-contraction
of $\JJM$ free fields is a fully well-defined notion.\footnote{It 
is believed that the conformal dimensions of chiral primary operators in four-dimensional $\mathcal N=2$
superconformal field theories are always rational.  This
is true automatically in Lagrangian theories and in
all known non-Lagrangian ${\cal N} = 2$ theories as well.  As
evidence in
the rank one case, all models in the general classification \cite{Argyres:2015ffa,Argyres:2015gha,Argyres:2016xmc,Argyres:2016xua} have rational conformal dimensions.}
However we may always perform a further transformation
to a logarithmic field ${\bf L} \equiv \log(\phi)$,
in which the calculation may be performed even for $\JJM
\notin \IZ$;  a similar point was made in \cite{Hellerman:2017veg}.  For the purpose of computing operator dimensions as
in \cite{Hellerman:2017veg}, the existence of the logarithmic
superfield is a convincing reason to believe there is no room for irregular
behavior depending on the fractional part of $\JJM$.  On
the other hand, the calculation of two-point functions is slightly different, as the non-single-valuedness of the
map may become relevant to the dynamics
at the point of insertion of the operator $\co$.  We will
leave this an open question, and for the present paper we will
simply choose $n$ such that $\JJM$ is an integer.  In
all Lagrangian theories, at any rate, $\D\ll\co$ is always
an integer, with the Coulomb branch chiral ring
being generated by traces of powers of the nonabelian
vector multiplet scalar.  For a rank-one Lagrangian theory,
$\D\ll\co$ is always $2$, and $\co$ lies in the multiplet of
the marginal operator cotangent to the microscopic
coupling $\t$. 

 \heading{Calculation of the free-field contribution}
 
 With all this in mind, we can now write the leading
 approximation to the two-point correlator.  Choosing
 $n$ so that $k = n\D$ is an integer and Wick contracting
 $\JJM$ complex free fields, we have
  \begin{align}\label{WickContractionAnswer}
\begin{split}
 \ampname\ll n \simeq \left< 
  \co\uu n \cob\vphantom{\co}\uu n\cc\right>\lrm{free}
 = { \mathcal P} \uu{\JJM}\cc (\JJM)!\ ,
\end{split}
\end{align}
 where
\begin{align}
\begin{split}
{\mathcal P}= (2\pi)\uu{-2}\cc
 \left|{\bf N}\ll\co\right|\sqd =
  {{g_{\text{eff}}^2}\over{4\pi\sqd}}
 \cc 
 \left|{\bf M}\ll\co\right|\sqd = {1\over{\pi\cc 
\Im(\t\lrm{eff})}}\cc
\left  |{\bf M}\ll\co\right|\sqd\ ,
\label{CalPCoefficientDef}
\end{split}
\end{align}
 where the $(2\pi)\uu{-2}$ comes from the normalization
 of the free propagator with unit kinetic term, \rr{PropagatorNormalizationForFieldWithUnitKineticTerm}.

This is simply the free approximation, of course, and
is not exact in $n$.  However in subsequent
sections we shall now show that interaction terms
have $n$-suppressed contributions in the quantum
effective action ${\cal B}\ll n$, 
using arguments parallel to those of \cite{Hellerman:2017veg} (also similar
arguments in nonsupersymmetric examples \cite{Alvarez-Gaume:2016vff,Loukas:2016ckj,Loukas:2017lof,
Hellerman:2015nra, Monin:2016jmo
,Hellerman:2017efx, Hellerman:2016hnf, Hellerman:2014cba,Hellerman:2013kba}). 

 In order to see this, we must relate our two-point function
 to a classical solution corresponding to the saddle
 point of the CFT action with sources corresponding
 to the logarithms of the operator insertions.  Doing this
 we shall check that the classical approximation
 to ${\cal B}\ll n$ matches
 \rr{WickContractionAnswer} up to terms of order
 $\log(\JJM)$ and smaller, which come from quantum
 effects in the path integral over the (free-field) action with (nonlinear) sources.

\subsection{Classical solution with operator insertions}

\heading{Large-$J$ insertions as classical sources}

The measure for the free Euclidean path integral
is of course ${\cal D}\phi \cc {\cal D}\phb\cc\eexp\left(-S\lrm{free}\right)$, so the path integral with insertions is equivalent to
a path integral with sources given by the negatives
of the logarithms of the insertions,
\begin{align}
\begin{gathered}
\left< \prod\ll i {\cal O}\ll i \right>\equiv \left< 1 \right>\uu{-1}\cc
\int \cc {\cal D} \phi \cc {\cal D}\phb \cc
 \eexp\left({- S\lrm {full}}\right)
= \int \cc {\cal D} \phi \cc {\cal D}\phb \cc \eexp\left(- (S\lrm {\rm free}
+ S\lrm{source})\right)\ ,
\\
S\lrm{source} \equiv - \sum\ll i\cc \log({\cal O}\ll i)\ .
\label{PathIntWithSources}
\end{gathered}
\end{align}
For us,
\begin{align}
\begin{split}
{\cal O}\ll 1 = [\co(x\ll 1)]\uu \JJM = ({\bf N}\ll\co)\uu \JJM\cc \phi\uu \JJM(x\ll 1)\ ,
\qquad
{\cal O}\ll 2 \equiv \left[\cob(x\ll 2)\right] \uu \JJM=\left( {\bf N}\ll\co\uu{*}\right)^\JJM \cc \phb\uu \JJM (x\ll 2)\ ,
\end{split}
\end{align}
where we have used \rr{CalOToPhiUnit} and defined 
\begin{align}\label{JNormalizationAndJToK}
\begin{split}
\JJM \equiv n\cc \D\ll\co = n |J\ll\co|\ ,
\end{split}
\end{align}
This quantity $\JJM \equiv n\D\ll\co$ is the $U(1)\ll R$ R-charge $J$ of the operator $\co\uu n$.

The full free classical action with sources, is then 
\begin{align}
\begin{gathered}
S\lrm{free+sources} =  - \JJM\cc \log(|{\bf N}\ll\co|\sqd)+\int \cc d\uu 4 x
\cc {\cal L}\lrm{dyn}\ ,
\\
 {\cal L}\lrm{dyn} \equiv \left(\pp_\mu\phb\right)\left(\pp_\mu\phi\right) - \JJM\cc \log(\phi)\d(x - x\ll 1) - \JJM\cc \log(\phb) \d(x - x\ll 2)  \ ,\label{DynLag}
\end{gathered}
\end{align}
so the EOM is
\begin{align}
\begin{split}
\pp\sqd\phb =  - {\JJM\over{\phi}} \cc \d(x - x\ll 1)\ ,
\qquad
\pp\sqd\phi  = - {\JJM\over{\phb}} \cc \d(x - x\ll 2)\ .
\end{split}\label{EOMWithSources}
\end{align}

\heading{Solution to the EOM}

The classical solution is not unique: It has a $U(1)_R$ phase zero mode but
no scaling zero mode.\footnote{The phase zero mode 
 only contributes a finite ($J$-independent) factor to the two-point function, as can
be seen from the free case.  This is not immediately
apparent because the solution is complex and 
the contour of integration for the phase zero mode
is not a priori obvious.  The correct contour of integration
and the finiteness of the phase zero mode integral, can be understood
more easily by organizing the calculation in terms of
coherent states and extracting Fock states from them,
which under the state-operator correspondence is
equivalent to adding linear rather than logarithmic 
sources and then performing a contour integral over
the strength of the linear source.  For our purposes all
this is irrelevant because the determinantal factors are
unchanged from the free case at the order of interest.}
The phase zero mode is an $R$-symmetry goldstone of the solution, which acts as
a constant shift of the axionic superpartner of the dilaton $\dilaton$.
An effective action
including this degree of freedom
 has been studied, and is
identified as the $\b$ field of \cite{Bobev:2013vta}.  At higher order in $\JJM$ the integral over the $\b$ zero mode
generates corrections to the quantum effective action through its measure,
but these are suppressed and only contribute at order $\JJM^0$ or smaller.  As
we are computing the quantum effective action only up to and including order
$\log(\JJM)$ in this article, we need not consider such effects.

So then the solution is of the form
\begin{align}
\begin{split}
\phi(x) = {{c\ll{\phi}}\over{(x - x\ll 2)\sqd}}\ , 
\qquad
\phb(x) = {{c\ll{\phb}}\over{(x - x\ll 1)\sqd}}\ ,
\end{split}\label{SolnParametrization}
\end{align}
and the equation of motion \rr{EOMWithSources}
is equivalent to
\begin{align}
\begin{split}
c\ll\phi \cc \pp\sqd \left [  (x - x\ll 2)\uu{-2} \right ]
= - {\JJM\over{\phb}} \d(x - x\ll 2)\ , 
\qquad
c\ll\phb \cc \pp\sqd \left [  (x - x\ll 1)\uu{-2} \right ]
= - {\JJM\over{\phi}} \d(x - x\ll 1)\ .
\end{split}
\end{align}
Using the normalization of the $\d$-function \rr{ScalarPropagatorNormalization}, this gives
\begin{align}
\begin{split}
- (2\pi)\sqd \cc c\ll\phi\cc
\d(x - x\ll 2) = - {\JJM\over{\phb}} \cc \d(x - x\ll 2)\ ,
\qquad
- (2\pi)\sqd \cc c\ll\phb\cc
\d(x - x\ll 1) = - {\JJM\over{\phi}} \cc \d(x - x\ll 1)\ ,
\end{split}
\end{align}
which is true if and only if
\begin{align}
\begin{split}\label{RelationOnStrengths}
(2\pi)\sqd \cc c\ll\phi\cc \phb(x\ll 2) =(2\pi)\sqd \cc c\ll\phb\cc \phi(x\ll 1) = \JJM\ . 
\end{split}
\end{align}
Plugging the parametrized solution \rr{SolnParametrization}
back into \rr{RelationOnStrengths} shows both
EOM are satisfied if and only if
\begin{align}
\begin{split}
(2\pi)\sqd \cc
{    { c\ll\phi c\ll\phb } \over{(x\ll 1 - x\ll 2)\sqd }}
 = \JJM
\end{split}
\end{align}
This is equivalent to
\begin{align}
\begin{split}
c\ll\phi = e^{i \b\ll 0} \cc c\ll 0,\qquad
c\ll\phb = e^{ - i \b\ll 0} \cc c\ll 0, 
\qquad
c\ll 0\equiv {{|x\ll 1 - x\ll 2|}\over{2\pi}}\cc \sqrt{\JJM}.
\end{split}
\end{align}
So the general solution is
\begin{align}
\begin{split}
\phi(x) &= {{c\ll \phi}\over{(x - x\ll 2)\sqd}} = {{e^{+ i \b\ll 0} \cc 
|x\ll 1 - x\ll 2|}\over{2\pi\cc (x - x\ll 2)\sqd}} \cc \sqrt{\JJM}\ ,
\\
\phb(x) &= {{c\ll \phb}\over{(x - x\ll 1)\sqd}} = {{e^{- i \b\ll 0} \cc 
|x\ll 1 - x\ll 2|}\over{2\pi\cc (x - x\ll 1)\sqd}} \cc \sqrt{\JJM}\ .
\end{split}\label{FinalSolnParametrization}
\end{align}
The magnitude of $\phi$ is
\begin{align}
\begin{split}
|\phi| \equiv (\phi\phb)\uu{\nicefrac12} = {{\sqrt{\JJM}}\over{2\pi}}\cc {{|x\ll 1 - x\ll 2|}\over{|x - x\ll 1|\cc |x - x\ll 2|}}\ , 
\end{split}\label{FinalSolnMagnitude}
\end{align}

\heading{Value of the action at the saddle point}

To evaluate the classical action, discard total derivatives
to rewrite the unit kinetic term as
\begin{align}
\begin{split}
|\pp\phi|\sqd = -\hh\cc \phi\cc\pp\sqd\phb -\hh\cc
\phb\cc\pp\sqd\phi\ , 
\end{split}
\end{align}
and the Lagrangian density then vanishes on the saddle point, except for delta-function contributions at the source
terms:
\begin{align}
\begin{split}
|\pp\phi(x)|\sqd = + {\JJM\over 2} \cc \d\upp 4(x - x\ll 1)
+ {\JJM\over 2} \cc \d\upp 4(x - x\ll 2)\ .
\end{split}
\end{align}
So the Lagrangian with sources \rr{DynLag}
can be reduced to 
\begin{align}
\begin{split}
{\cal L}\lrm{dyn} \simeq - \JJM\cc \left ( \log(\phi) + \hh  \right)\cc \d\upp 4(x - x\ll 1)
- \JJM\cc
\left ( \log(\phb) + \hh  \right)
\cc \d\upp 4(x - x\ll 2)
\end{split}
\end{align}
where the $\simeq$ indicates that we have discarded
total derivatives.
The action of the classical solution at the saddle point
is therefore
\begin{align}
\begin{split}
S\lrm{dyn} = \int \cc d\uu 4 x \cc {\cal L}\lrm{dyn} = - \JJM
\cc \log
\left|\phi(x\ll 1)\phb(x\ll 2)  \right|
 + \JJM\ .
\end{split}
\end{align}
Plugging in the solution \rr{FinalSolnParametrization},
we have
\begin{align}
\begin{split}
\left|\phi(x\ll 1)\phb(x\ll 2)\right| = 
{\JJM\over{(2\pi)\sqd\cc|x\ll 1 - x\ll 2|\sqd}},
\end{split}
\end{align}
which gives
\begin{align}
\begin{split}\label{SaddlePointDynamicalAction}
S\lrm{dyn} = \JJM\cc\left [ \cc - \log(\JJM) + 1 + 2\cc \log|x\ll 1 - x\ll 2| + 2\cc\log(2\pi) \cc \right ]\ .
\end{split}
\end{align}
at the saddle point.

\heading{Classical approximation to the free two-point function}

We see that the total classical action goes as $\JJM$ (with a crucial logarithmic enhancement in the source contribution), where from \rr{JNormalizationAndJToK} $\JJM$ is the total $R$-charge of the operator $\co\uu n$.  Thus the total $R$-charge $J$ of the operator acts as an inverse Planck constant $\hbar\uu{-1}$,
and suppresses quantum fluctuations of any operator product $\prod\limits_i \co\ll i\uprm{other}$ insterted into
the two-point function.  In particular, the field $\phi$ in the nonlinear source term, could be divided into their classical value
plus fluctuation piece, $\phi = \phi\lrm{cl} + \phi\lrm{fluc}$, and $ \JJM$ acts as a parameter suppressing nonlinear
quantum effects relative to the classical partition function $Z\lrm{cl} \equiv \eexp({ - S\lrm{free+sources}})$.  That is,
we expect 
\begin{align}
\begin{split}
\log(Z\lrm{free+sources}) \simeq\log(Z_{\text{free$+$sources, classical}}) = - S\lrm{free+sources}\ ,
\end{split}
\end{align}
with errors of relative order $\JJM\uu{-1}$.

 Let us check this explicitly, to verify that $J$ really does act as a quantum loop-suppressing parameter.  Since the
 exact classical partition function $Z\lrm{free+sources}$ is given exactly by the Wick contraction formula, 
 we only have to compute the saddle-point value of the classical free action with sources, and compare it
 to the asymptotic expansion of the logarithm of the Wick-contraction result \rr{WickContractionAnswer}.

Adding the constant piece $- k\JJM\cc  \log(|{\bf N}\ll\co|\sqd) $ to the dynamical action, we have the value of the full free
action with sources \rr{DynLag} at the saddle point:
\begin{align}
\begin{split}\label{FullFreeSaddlePointAction}
S\ll {{\text{free} + \text{sources}}} =  - \JJM \cc  \log(|{\bf N}\ll\co|\sqd) + S\lrm{dyn}\ ,
\end{split}
\end{align}
with $S\lrm{dyn}$ given in \rr{FullFreeSaddlePointAction}.

Defining $Z\ll n$ to be the full CFT path integral
with sources,
then
\begin{align}
\begin{gathered}
Z\ll 0\uu{-1}\cc Z\ll n \simeq
\eexp\left(
-
S\ll {{\text{free} + \text{sources}}} 
\right) = (2\pi)\uu{-2\JJM}\cc |x\ll 1 - x\ll 2|\uu{-2\JJM}\cc 
|{\bf N}\ll\co|\uu{2\JJM}\cc \JJM\uu \JJM\cc e^{-\JJM}\ ,
\end{gathered}
\end{align}
Then using our definition \rr{TwoPtAmpDef},
\rr{RescalingAbbrevDef} of the rescaled two-point function and the result \rr{WickContractionAnswer},
we have
\begin{align}
\begin{gathered}\label{ClassicalFreeApproxToTwoPtFunc}
\ampname\ll n \simeq (2\pi)\uu{-2\JJM}\cc |{\bf N}\ll\co|\uu{2\JJM}\cc \JJM\uu \JJM\cc e^{-\JJM}\ ,
\qquad
\JJM = n\D\ll\co\ .
\end{gathered}
\end{align}
in the classical approximation to free field theory.
This approximation
can be interpreted as the normalization factor $|{\bf N}\ll\co|\uu {2\JJM}$, times Stirling's approximation to the 
Wick-contraction $ (2\pi)\uu{-2\JJM}\cc \cc \JJM!$ of $\JJM = n\D\ll\co$ free fields separated by unit distance.

So we have verified that the total $R$-charge $\JJM$ really is acting as a
loop-suppressing parameter, as expected.  
We will now see that this is a useful point of view for bounding the size of
subleading corrections to ${\cal B}\ll n$ at large $\JJM$.
If we intended to stop at this level of accuracy, order $\JJM$ and $\JJM\cc\log (\JJM)$,
the saddle point estimate \rr{ClassicalFreeApproxToTwoPtFunc} would be a rather clumsy way to approximate
a free-field correlation function; if the action were exactly free,
then we would just use the more accurate exact
formula \rr{WickContractionAnswer}.
However the estimate of
the large-$n$ Wick contraction by a classical
saddle point, makes it possible to go beyond
free-field approximations, and include the effects
of interaction terms.  In the next section we turn to
the inclusion of interaction terms, in particular searching
for interaction terms that make contributions to ${\cal B}\ll n
= \log(\ampname\ll n)$ that are larger than order $\JJM\uu 0$.

\section{Contributions of interaction terms}

In equation \rr{ClassicalFreeApproxToTwoPtFunc}
we have written an estimate for the two-point function
$\ampname\ll n$, with the symbol $\simeq$ indicating
that we have ignored terms beyond
the free-field action on moduli space.  We would like
to know how accurate an approximation that $\simeq$
actually represents.  In order to do so, we must
estimate the size of interaction terms at large $n$.
Since $n$ controls the size of the vev of $|\phi|$ in the
classical solution, with the magnitude of
$|\phi|$ going as $|\phi|\propto \sqrt{\JJM}$,
we can estimate the $n$-scaling or equivalently the $\JJM$-scaling of the leading contribution of an individual term in the action, by estimating its $|\phi|$-scaling.  

\subsection{$J$-scalings and the dressing rule}

By the $|\phi|$-scaling of a term in the effective
action on moduli space, we mean simply
the number of $\phi$'s and $\bar{\phi}$'s appearing
in the numerator of the term, minus the number
of $\phi$'s and $\bar{\phi}$'s appearing
in the denominator.  In the denominator, the fields can
only appear undifferentiated. This rule, long (correctly)
treated as self-evident for study of moduli space
effective actions, has its origin in the nontrivial fact that moduli spaces exist, and so
the leading term in the denominator in an effective action for moduli, must be an undifferentiated field,
in contrast to cases such as
\cite{Aharony:2009gg,Aharony:2011ga,Aharony:2013ipa,Alvarez-Gaume:2016vff,Hellerman:2014cba,Hellerman:2015nra,Monin:2016jmo}
in which the dressing for singular terms is a differentiated field.

 More generally, in any
effective field theory spontaneously breaking scale
invariance, the dressing field appearing in the denominator
of a term must be a dimensionful field which has
an expectation in the spontaneously broken vacuum,
and functions as an (exponentiated) effective dilaton.

The spontaneously broken scale invariance and
$R$-symmetry give quite
powerful selection rules on terms that can appear
in the effective action, even without using the full power
of superconformal invariance.  The most important
point is that, since $\phi$ and $\phb$ themselves are the
lowest-dimension fields with vevs, they are the
unique dressing fields to render a term
scale invariant and $R$-symmetry invariant.  For instance,
if $({\mathsf{term}})\lrm{undressed}$ is some
monomial in $\pp\uu k\phi, \pp\uu k\phb$, the abelian
gauge field strength ${\cal F}\ll{\m\n}$, and its derivatives,
and the fermions $\psi,\psb$ in the vector multiplet
and their derivatives, then the term has a unique
scale-invariant and $U(1)$ $R$-symmetry-invariant dressing
by $\phi$ and $\phb$:
\begin{align}
\begin{gathered}
({\mathsf{term}})\lrm{dressed} = \phi\uu{-\ell}\cc \phb\uu{-\tilde{\ell}} ({\mathsf{term}})\lrm{undressed}\ ,
\\
\ell = \hh\cc\big (\cc \D\lrm{undressed} +  J\lrm{undressed}\cc \big )\ ,
\qquad
\bar{\ell} = \hh\cc\big (\cc \D\lrm{undressed} -  J\lrm{undressed}\cc \big )\ ,
\end{gathered}
\end{align}
where $\D\lrm{undressed}$ and $J\lrm{undressed}$ 
are the dimension and $R$-charge of the undressed term.

 The latter can be seen more clearly 
by organizing terms into superspace integrals,
over all the Grassman variables (${\cal N} = 2$ $D$-terms) or a subset (${\cal N} = 2$ $F$-terms and $\th\uu 6$-integrals).
Then the term is guaranteed to be supersymmetric 
assuming the $F$-term or $\th\uu 6$-integrand
satisfies the appropriate restriction of being invariant
already under the SUSYs corresponding to the unintegrated
Grassman variables.  The dressing rule can then be
implemented at the level of the superspace integrands
themselves, taking into account the contributions
of the superspace measure to the conformal
dimension and $R$-charge of the operator. 

Representing the effective vector multiplet
as
 the 
superfield $\Phi$ whose lowest component is $\phi$,
the dressing rules translate into a the need
to dress each derivative $\pp\ll\m$ or $D\ll\a\uu i, \bar{D}\ll\ald\uu i$ to be scale-invariant and $U(1)\ll R$-invariant,
using undifferentiated $\Phi$'s and $\bar\Phi$'s
themselves.  Each $\pp$ must be compensated
by a dressing $(\Phi\bar\Phi)\uu{-\hh}$ and
each $D\ll\a\uu i$ or $\bar{D}\ll\a\uu i$ must be cancelled
by a $\Phi\uu{-\nicefrac12}$ or $\bar\Phi\uu{-\nicefrac12}$, 
respectively.  Since it is the total number of $\Phi$
and $\bar\Phi$ that determine the overall $J$-scaling
of a term, it is the dimension of the undressed
term that determines the overall $J$-scaling of the
dressed term, with $U(1)\ll R$-symmetry invariance
entering as an additional condition determining the
individual number of $\Phi$'s and $\bar{\Phi}$'s.

Before starting to classify terms, we mention several caveats for the reader to bear in mind:
\bi
\item{Dressing with the correct number of $\Phi$ and
$\bar\Phi$ in the denominator, is necessary but obviously far from a sufficient criterion for
a term that can appear in the effective action: Apart
from supersymmetry and rigid scale invariance, we are
ignoring many other symmetries such as $SU(2)$ R-symmetry and also
Weyl-covariance on curved backgrounds.  These restrict terms very severely, but 
as we shall see supersymmetry, $U(1)$ R-symmetry,
and scale invariance alone eliminate all possible contributions from superconformal interaction
terms up to order $\JJM\uu 0$. }
\item{Our classification of terms is done in flat space; we
should be careful because this classification can underestimate the $\JJM$-scaling of a term in a general curved
background.  There can be super-Weyl-invariant
terms involving curvatures, where the $J$-scaling of the
curvature-dependent piece is actually larger than the
$J$-scaling of the Weyl-completion in flat space.  One
such example occurs even in the simple example of the
large-charge effective theory of the $O(2)$ model,
in which the Weyl-invariant term 
\begin{align}
\begin{split}
|\pp\chi|\cc {\tt Ric}\ll 3+
2 {{(\pp|\pp\chi|)\sqd}\over{|\pp\chi|}}
\end{split}
\end{align}
 has a curvature-dependent piece scaling as $\JJM\uu\hh$, and a Weyl-completion scaling only as $\JJM\uu{-\nicefrac12}$
 for a helical solution
  in flat space.}
\item{To classify terms systematically including their Weyl-invariant curvature completions, one ought
to use a superconformal generalization of the formalism of \cite{Monin:2016jmo}.  This beautiful formalism can be used in large-$J$ effective theories such as \cite{Hellerman:2017veg,Hellerman:2015nra, Alvarez-Gaume:2016vff, Monin:2016jmo, Loukas:2016ckj,Loukas:2017lof} to classify terms efficiently in goldstone boson actions when nontrivial combinations of global charge and conformal invariance can be preserved with the rest spontaneously broken.\footnote{While expanding the previous scope of the CCWZ formalism\cite{Coleman:1969sm,Callan:1969sn}, the formalism of \cite{Monin:2016jmo} still requires a translationally invariant ground state to 
be applied straightforwardly.  It would be quite interesting
to understand how to generalize the method of \cite{Monin:2016jmo} to cases in which the unbroken
symmetry group does not act transitively on spacetime
events, as in such examples as \cite{Hellerman:2013kba,Hellerman:2017efx}, or the separate
case of a conformal striped phase, in which the inhomogeneity would be at the scale defined by the charge density.} For ${\cal N} = 2$ superconformal theories in $D=3$, for instance, a suitably adapted conformal supergravity formalism such as \cite{Kuzenko:2015jda} would be tantamount to a superconformal extension of \cite{Monin:2016jmo}
and indeed the formalism of \cite{Kuzenko:2015jda}
was used in \cite{Hellerman:2017veg}
to construct the first subleading large$-J$ correction
to the moduli space effective action.}
\item{However in the present
effective theory it is easy to see that the only
possible scale- and $U(1)\ll R$-invariant curvature-dependent term scaling as greater than $\JJM\uu 0$, is the 
term $(\tt Ric)\ll 4 \cc |\phi\lrm{unit}|\sqd$.  This curvature-dependent term is simply
the usual conformal coupling that Weyl-completes the
flat-space kinetic term $|\pp\phi\lrm{unit}|\sqd$, with
coefficient $\nicefrac16$.  Adding even one more curvature 
would require an additional $|\phi|\uu{-2}$ in the dressing
if the curvature invariant is a scalar.  Nonscalar curvature terms are not dangerous either:  The upper Ricci tensor,
for instance, has weight $+4$ under a rigid Weyl rescaling, and would have to
be cancelled with $(\pp\ll\m \phi \pp\ll\n\phb) / |\phi|\sqd$
in order to have weight $4$, making the total
$\JJM$-scaling vanish.  Curvature invariants
with more free indices, have even higher
Weyl weight, and need even more undifferentiated
$|\phi|$'s to make a term of total weight $4$ after
contracting with derivatives of the fields.  We will
organize our search for terms, therefore, around
superconformal terms in the flat-space Lagrangian.}

\item{This procedure for classifying terms applies only
to superconformal terms in the Wilsonian action, \it i.e. \rm 
terms that are themselves invariant under the Weyl
and $R$-symmetries.  Since the underlying CFT is invariant
modulo the anomaly, so must the effective action
be invariant, modulo the underlying anomaly, which
is
 partially expressed by Wess-Zumino terms compensating the anomaly between the underlying CFT and the effective
theory of moduli space.  The coefficients of these
terms are $c$-numbers, independent of the state, 
so the coefficient cannot have any $J$-dependence
and the only independent Wess-Zumino term occurs
at order $J\uu 0$ in flat space, with an enhancement
to $\log (J)$ in the presence of curvature.}
\ei
\subsection{Dressing and $J$-scaling of superconformal interaction terms}\label{HigherDerivConformal}

Now we shall consider superconformal interaction terms.  We shall
show that no superconformal interaction term
can give a contribution of order $n\uu 0$ or larger
in ${\cal B}\ll n$ at classical level.  Since loop contributions of any such terms must be even further suppressed by powers
of $n$, we will know that we can obtain an estimate
for ${\cal B}\ll n$ accurate up to order $n\uu 0$ if
we exclude interaction corrections appearing as
superconformal terms, leaving only the Wess--Zumino
term as a candidate contribution larger than $n\uu 0$.

\heading{Higher-derivative terms for vector multiplets}

As noted in the introduction, quantum and classical effects coming from  ${\cal N} = 2$ superconformal higher-derivative operators,
are no larger than order $n\uu 0$.  The analysis is parallel with that in \cite{Hellerman:2017veg}:
each $D\ll\a$ or $\bar{D}\ll\a$ derivative in a superspace action must be dressed with at least one $\phi\uu{-\hh}$ or
$\phb\uu{-\hh}$, and each $\pp_\mu$ must be dressed with at least one $|\phi|\uu{-1}$, in order to preserve scale invariance.
The full-superspace measrure $d\uu 8 \th$ has dimension $4$, and so a full superspace integrand must have
\begin{align}\label{JScalingCountingDTm}
    \begin{alignedat}{99}
 {\text{$D$-term dressing:}}  &&\qquad\cc N\ll{\phi + \phb} &= - N\ll{\pp} - \hh\cc N\ll{D + \bar{D}}
\ ,
\\
{\text{$D$-term $J$-scaling:}}& &\qquad \hh \cc N\ll{\phi + \phb}   &= - \hh N\ll{\pp} - {1\over 4}\cc N\ll{D + \bar{D}}
\ .
\end{alignedat}
\end{align}
Therefore, even a full-superspace integrand with
no derivatives would have to have scaling $J\uu 0$.  The only superconformal full-superspace integrand would
therefore be the identity, which vanishes when integrated over superspace.  Any nonvanishing full-superspace integral
must have negative $J$-scaling.   

A half-superspace integrand ($F$-term) has dimension two, so 
\begin{align}\label{JScalingCounting}
    \begin{alignedat}{99}
 {\text{$F$-term dressing:}}  &&\qquad N\ll{\phi + \phb} &=  2 - N\ll{\pp} - \hh\cc N\ll{D + \bar{D}}
\ ,
\\
{\text{$F$-term $J$-scaling:}}&& \qquad \hh\cc N\ll{\phi + \phb} &= 1 - \hh\cc N\ll{\pp} - {1\over 4}\cc N\ll{D + \bar{D}}\ .
\end{alignedat}
\end{align}
The $F$-term integrand with zero derivatives is just the classical kinetic $F$-term proportional to $\phi\sqd$.
The next term has been found  \cite{Argyres:2008rna, Argyres:2003tg, Argyres:2004kg, Argyres:2004yp} to
have $N\ll{\pp} = 2$ as a half-superspace integrand, given in equation (5.13) of \cite{Argyres:2003tg}.
The integrand is proportional to ${\cal G}[\phi](\pp\phi)(\pp\phi)$ for some holomorphic function ${\cal G}(\phi)$.
This integrand must have dimension $2$, so the only allowed term is ${\cal G} [\phi]= \phi\uu{-2}$.  However
the $R$-charge of any $F$-term integrand must be the same as that of $\phi\sqd$, namely $J\ll F =  2$ in our $R$-charge convention,
whereas the $R$-charge of $\phi\uu{-2}\cc (\pp\phi)(\pp\phi)$ vanishes, and it is therefore not an admissible term.  It follows that
any contributing superconformal $F$-term in the Coulomb branch effective theory, must have strictly negative $J$-scaling. 

\heading{Inclusion of neutral massless hypers}

The above argument holds for pure Coulomb branches, with no massless hypermultiplets present.  If there are massless hypermultiplets,
the analysis is slightly subtler, because there can be higher-derivative effective terms
containing both vector and hypermultiplet degrees of freedom.  Such terms can even come in the form
of integrals over six of the eight supercharges.  Such a $\th\uu 6$ integrand has 
\begin{align}\label{JScalingCountingThreeQtrsSS}
    \begin{alignedat}{99}
 {\text{$\theta^6$-term dressing:}}  &&\qquad N\ll{\phi + \phb} &= 1 - N\ll{\pp} - \hh\cc N\ll{D + \bar{D}} - N\ll h
\ ,
\\ 
{\text{$\theta^6$-term $J$-scaling:}}&& \qquad 
\frac12N\ll{\phi + \phb} &= \hh - \hh\cc N\ll{\pp} - {1\over 4}\cc N\ll{D + \bar{D}} - \hh\cc N\ll h\ .
\end{alignedat}
\end{align}
where $N\ll h$ is the number of powers of hypermultiplets appearing in the term.

  The classical solution has a VEV only for the vector multiplet scalar, because
the classical solution involves only sources for the vector modulus $\phi\lrm{unit}$, and the moduli space metric
factorizes between vector multiplet and hypermultiplet factors.  In the classical solution, all degrees of freedom in the hypermultiplets
are set to zero and so the only terms that can contribute with the maximal $\JJM$-scaling of the term, are those containing only vector multiplet degrees of freedom.  In other words, the only terms that contribute classically are those with $N\ll h = 0$,
and these can have $\JJM$-scaling at most $\JJM^{\nicefrac 12}$.  
However any
term in the component action containing only vector multiplets,
would be supersymmetric on its own, and would correspond to one of the pure vector-multiplet/pure Coulomb-branch terms in \cite{Argyres:2003tg}.
And according to the results of \cite{Argyres:2003tg}, there are no $\th\uu 6$ terms whose integrands
have $N\ll\pp + \hh N\ll{D + \bar{D}}  = 1$, only ordinary $F$- and $D$-terms.  We have already considered such 
terms in the previous discussion, and
shown that they  contribute only smaller than order $\JJM\uu 0$.  Mixed vector-hyper terms can appear, of course, 
but their classical value is zero and they only contribute 
through their quantum effects.  Each quantum loop gives an additional suppression of $1/\JJM$, relative to the maximum $\JJM$-scaling given by eq \rr{JScalingCountingThreeQtrsSS}.   This can be thought of as coming from the need to Wick-contract at
least two hypermultiplet degrees of freedom in the term.  So any mixed-branch terms in \cite{Argyres:2003tg} which are
not pure vector-multiplet terms, must have $N\ll h$ at least $1$, and therefore negative $\JJM$-scaling.

\heading{Order $\log(J)$ term from Wess--Zumino coupling}

The analysis above would seem to rule out any possible effect larger than $O(J\uu 0)$, with only the
determinant in the free theory contributing at that order.  This is not 
quite
accurate however, as we have
so far considered only terms that are manifestly superconformal terms in a Wilsonian effective action.
This is not quite the case, however:   There is a unique
interaction term in the effective Lagrangian that does not correspond
to a superconformal term, namely the
four-derivative Wess--Zumino term in the bosonic action \cite{Komargodski:2011vj,Schwimmer:2010za} and its supersymmetric completion.
This term makes
contributions of order $\log(\JJM)$ and $\JJM\uu 0$.
This is the only term, therefore, that can contribute
a power-law factor $\JJM\uu\a$ to the amplitude $\ampname\ll n$,
and its coefficient $\alpha$ is determined by the coefficient
of the $a$-anomaly.  Since we are only computing
$\log(\ampname\ll n)$ up to the order $\log(\JJM)$,
the Wess--Zumino term is the only interaction term we will
need to consider, and only its classical effect will be important, with its quantum effect contributing to
$\log(\ampname\ll n)$ at order $\JJM\uu{-1}$ and smaller.

The only $\log(\JJM)$ piece of the 
Wess--Zumino term is in the coupling
of the dynamical effective dilaton to
 to the Euler
density of the background metric; all other terms contribute strictly $\JJM\uu 0$ classically, and smaller quantum-mechanically.
 This term is invisible in the $\IR\uu 4$
conformal frame and we must transform
to the $\Sfour$ frame in order to see it clearly.  We
shall explain why this is necessary this in section \ref{SphereSec}.

\subsection{Structure of the asymptotic expansion at large $n$}\label{AsympSumRules}
Before evaluating the Wess--Zumino contribution at
order $\log(\JJM)$, we would like to express the 
structure of the asymptotic expansion of 
${\cal B}\ll n = \log(\ampname\ll n)$ as we have understood
it so far.

We now have all the information we need to write
the leading terms in the asymptotic expansion of
${\cal B}\ll n$ up to order $\log(\JJM)$.  The amplitude
$\ampname\ll n$ can be thought of as a partition function 
with sources $- n\cc \log(\co)$, normalized by the
partition function without sources, which is just the
sphere partition function:
\begin{align}
\begin{split}
\ampname\ll n = \eexp\left({{\cal B}\ll n}\right) = |x - y|\uu{2\JJM}\cc
Z\ll 0\uu{-1} \cc Z\ll n\ ,
\end{split}
\end{align}
where 
$Z\ll n$ is the path integral 
with 
integrand $\eexp({- S\ll n\uprm{full}})$, 
\begin{align}
\begin{split}\label{Snfull}
S\ll n\uprm{full} \equiv S\lrm{CFT} - n \cc \log(\co(x))
- n \cc \log(\co(y))\ .
\end{split}
\end{align}
At large $n$, the path integral is dominated by the saddle
point described by the classical solution \rr{FinalSolnParametrization} in which
$\langle \phi \rangle$ is large and conformal invariance is
spontaneously broken.  In this regime we approximate
$S\lrm{CFT}$ by its moduli space effective action
and identify $\co$ with ${\bf N}\ll\co \cc\phi\lrm{unit}$.
Then the quantity ${\cal B}\ll n$ is simply the difference
in the quantum effective action with sources $n$
from that with vanishing sources:
\begin{align}
\begin{split}
{\cal B}\ll n = - \log(Z\ll n) + \log(Z\ll 0)
+ 2 \JJM\cc \cc\log|x - y|\ .
\end{split}
\end{align}

From this point of view, it is natural that ${\cal B}\ll n$ should
have a well-behaved ${1/ n}$ expansion
at large $n$, since it is a sum of connected Feynman
diagrams in a path integral whose action is proportional
to $\JJM = n\D\ll\co$.  Indeed the only surprise is that there should
be any terms nonanalytic in $n$ at all.  The nonanalytic terms cannot arise from infrared-singular dynamics
of the effective theory, for the effective theory is infrared-free.
The origin of the nonanalytic terms going as $\log(n)$
and $n\cc \log(n)$ is more banal: They appear
because of the explicit nonanalyticity of the source
term as a function of $\phi$: Since $\phi$ scales as
$\JJM\uu{\nicefrac 12}$ and there is an explicit $n\cc\D\ll\co \log(\phi)$ 
term in the action $S\ll n\uprm{full}$, there is
a classical $n\cc \log(n)$ term as
well as a $\log(n)$
term, where the latter can be thought of as a one-loop effect in quantizing around the classical saddle point, or more
efficiently as a subleading term in the large-$n$ expansion
of $\log[ (n\D_{\mathcal O})!]$ by Stirling's formula.  

In the previous sections we have shown that the explicit
superconformal interaction terms never contribute larger
than $n\uu 0$ in the quantum effective action even
through their classical value; their quantum effects are even smaller.  The only term making a contribution larger 
than $n\uu 0$ is the Wess--Zumino term, which
contributes at order $\log(n)$ with a coefficient
is proportional to an anomaly mismatch $\Delta a \equiv a\lrm{CFT} - a\lrm{EFT}$.  

So altogether, up to order $n\uu 0$ terms, we have
\begin{align}
\begin{split}
{\cal B}\ll n = \log[(\JJM)!] + b\ll{-1}\cc \JJM + \a\cc
\log(\JJM) + O(n\uu 0)\ ,
\qquad
b\ll {-1} \equiv  2\cc \log\left ( {{|{\bf N}\ll\co|}\over{2\pi}}
\right )\ ,
\end{split}
\end{align}
and $\a$ determined by the conformal
anomaly via the Wess--Zumino term.  
Exponentiating,
\begin{align}
\begin{split}
\ampname\ll n = 
\JJM! \cc \left ({{|{\bf N}\ll\co|}\over{2\pi}} \right)\uu {2\JJM} \cc  (\JJM)\uu\a \cc  \tilde{\ampname}\ll n\ ,
\end{split}\label{MainFormulaForAmpname}
\end{align}
where  $\tilde{\ampname}\ll n$ approaches a constant as $n\to\infty$, and ${\bf N}\ll\co$ is an $n$-independent constant describing
on the normalization of the operator relative to the effective Abelian
gauge coupling.
The exponent $\a$ is a positive
number proportional to the difference between the $a$-anomaly coefficient of the underlying CFT and that
of the effective theory of the Coulomb branch.
In section \ref{AnomalySec}, we will calculate of the coefficient $\a$ in terms of the conformal $a$-anomaly.

\newpage

\heading{Sum and product rules}

Despite the simplifications of ${\cal N} = 2$ superconformal
symmetry for chiral primary correlation functions, the
computation of two-point functions is still nontrivial
and not solved except in some simple cases.
For some theories, we may be able to obtain only numerical
or approximate data, against which we may want
to check our predictions at large $n$.
In such cases, it is useful to express the properties of
the asymptotic expansion in the form of sum rules for
${\cal B}$, or equivalently product/quotient rules
for $\ampname\ll n$, which isolate particular terms in
the asymptotic expansion combining adjacent terms
at large $n$.  

The simplest rules are simply limits for ${\cal B}\ll n$,
\begin{align}
\begin{split}
 {{{\cal B}\ll n}\over{\log [(\JJM)!]}} = 
1 + O\left({1\over{\log (\JJM)}}\right)\ .
\end{split}\label{SumRuleA}
\end{align}
The error comes from the operator-normalization-dependent term $b\ll{-1}
\JJM$.  The inverse of a logarithm falls off very slowly, so it would be better not to have this term present.  Of course, if we already know the normalization ${\bf N}\ll\co$ we could
write the more precise limit
\begin{align}
\begin{split}
{{{\cal B}\ll n -    b\ll{-1}\cc\JJM   }\over{\log [(\JJM)!]}}   = 1 + O( n^{-1})\ .
\end{split}\label{SumRuleB}
\end{align}
However it is cumbersome to extract the normalization of $b\ll {-1} = \log [{\bf N}\ll\co / (2\pi) ]$: This coefficient depends on
the normalization of the operator $\co$ itself and
so combines information from many choices of conventions that are not straightforward to compare among definitions of $\co$.  It is more convenient
to find sum rules that cancel 
the factor ${\bf N}\ll\co$, so that we do not have to bother
matching it with any other normalization.

The most straightforward such sum rule is
\begin{align}
\begin{split}
n   {\cal B}\ll{n+1} 
- (n + 1)   {\cal B}\ll n  =
\JJM -\left (\a + \hh\right) \log \left (  \JJM \right)  + O\left(n\uu 0\right) \ .
\end{split}\label{SumRuleC}
\end{align}
This version of the sum rule looks particularly stringent, because
the individual terms on the LHS scale as $n\sqd\cc\log (n)$ while the error on the LHS scales as $n\uu 0$,
and yet $|{\bf N}\ll\co|$ drops out of the rule
completely, making it convenient to check.

Extracting the logarithm may be cumbersome analytically
or costly computationally, so it may in some cases
be better to check product rules rather than
sum rules.  Exponentiating the sum rule \rr{SumRuleC}
directly gives
\begin{align}
\begin{split}
{{ \left ( \ampname\ll {n+1} \right )\uu n }\over{  
\left ( \ampname\ll n \right )\uu {n+1}}   } = {{\eexp\left({\JJM}\right)}\over{(\JJM)\uu{\a + \hh}}} \cdot
\left [  O(n\uu 0) + O(n\uu{-1}) + \cdots \right ]\ .
\end{split}\label{SumRuleD}
\end{align}

The difficulty of this sum rule is that it involves
raising numbers of order $(\JJM)!$ to the $(n+1)\uu{\underline{\rm th}}$ power and taking
ratios of them; both numerator and denominator have
of order $n\sqd\cc\log (n)$ digits, which cancel with a precision
of $O(n)$ digits, so a great many significant figures of precision are wasted.

A product rule or corresponding sum rule can evade
this difficulty and still cancel the normalization factor
${\bf N}\ll\co$, by using three adjacent neighboring values,
raised only to the power $1$ or $-2$
in the product rule, so that both numerator and
denominator of the product have only $O(n)$ digits each.
We have:
\begin{align}
\begin{split}
{{\ampname\ll{n+2}\cc\ampname\ll n}\over{\ampname\ll{n+1}\sqd}}=  \eexp\left({  {{\D\ll\co}\over n} - {{\D\ll\co + \hh + \a}\over{n\sqd}}}
\right)
\left [ \cc 1 + O(n\uu{-3}) \right ]\ ,
\end{split}\label{SumRuleE}
\end{align}
equivalent to a sum rule for ${\cal B}\ll n$ approximating
a discretized second derivative,
\begin{align}
\begin{split}\label{SumRuleF}
{\cal B}\ll {n + 2} + {\cal B}\ll n - 2 \cc {\cal B}\ll{n+1}
= {{\D\ll\co}\over n} - \frac{1}{n^2}\left(\D\ll\co + \hh + \a\right)
+ O(n\uu {-3})\ .
\end{split}
\end{align}
This rule, \rr{SumRuleF}, is somehow the most convenient
expression of the asymptotic expansion for verifying the
formula, because it allows us to perform
three independent consistency checks, at orders $n\uu{0,-1,-2}$, respectively, without great computational
difficulty given the coefficients
${\cal B}\ll n$, and does not require
knowledge of the normalization $|{\bf N}\ll\co|$.  
The independent checks can be expressed as:
\begin{align}
\label{SQCDSumRuleFOrd0}
\lim_{n\to\infty}
\left(
{\cal B}\ll {n + 2} + {\cal B}\ll n - 2 \cc {\cal B}\ll{n+1} 
\right)
&= 0\ ,
\\
\label{SQCDSumRuleFOrd1}
\lim_{n\to\infty} \cc n \left (  
{\cal B}\ll {n + 2} + {\cal B}\ll n - 2   {\cal B}\ll{n+1}
  \right ) 
  &= \D\ll\co\ ,
\\
\label{SQCDSumRuleFOrd2}
\lim_{n\to\infty}    \left [
n\sqd\left( 
{\cal B}\ll {n + 2} + {\cal B}\ll n - 2 \cc {\cal B}\ll{n+1} \right)
 - n\cc\D\ll\co  \right]
 &= - \left(\D\ll\co + \hh + \a \right)\ . 
\end{align}
The multiplicative version
of the rule is
\begin{align}
\begin{split}\label{SQCDSumRuleG}
{{\ampname\ll{n+2}
\ampname\ll n}\over{\ampname\ll{n+1}\sqd}} = 1 + {{\D\ll\co}\over n} + \hh\left(  \D\ll\co\sqd - 2\D\ll\co - 1 - 2\a  \right )  n\uu{-2} + O(n\uu{-3})\ ,
\end{split}
\end{align}
whose individual components are
\begin{align}
\label{SQCDSumRuleGOrd0}
\lim_{n\to\infty}  {{\ampname\ll{n+2}\ampname\ll n}\over{\ampname\ll{n+1}\sqd}} 
&= 1 \ , 
\\
\label{SQCDSumRuleGOrd1}
\lim_{n\to\infty} n \left (  {{\ampname\ll{n+2}\ampname\ll n}\over{\ampname\ll{n+1}\sqd}}  - 1 \right)
&= \D\ll\co\ ,
\\
\label{SQCDSumRuleGOrd2}
\lim_{n\to\infty}   n\sqd \left(  {{\ampname\ll{n+2}\ampname\ll n}\over{\ampname\ll{n+1}\sqd}}  - 1  - {{\D\ll\co}\over n} \right) 
&= \hh \left ( \D\ll\co\sqd - 2\D\ll\co - 1 - 2\a  \right ) \ .
\end{align}
In  section \ref{localization} we shall check \rr{SumRuleF} and \rr{SQCDSumRuleG} 
in two cases
 with  $\D\ll\co = 2$
 using
results from supersymmetric localization, after
computing the value of $\a$ in terms of the conformal
$a$-anomaly.

\subsection{Correlators on $\IR\uu 4$ vs. $\Sfour$}
\label{SphereSec}

The evaluation of the $\log(n)$ contribution to the quantum effective action, will be the main
nontrivial part of the effective field theory calculation.
Our next step should be to Weyl-transform the solution to
the sphere $\Sfour$, because the $\log(n)$ term
is invisible in flat space. 

\heading{Why should we need to consider $\Sfour$ at all?}

 Before doing so, though, we should
explain briefly why one must consider the $S\uu 4$ conformal frame at all.  After all,
our basic paradigm is to quantize the effective theory
in the background of the classical solution, and this
should in principle work equally well on $\IR\uu 4$ as
on $\Sfour$.  In other words, if the $\log(\JJM)$ term
is invisible on $\IR\uu 4$, then where is it and why
can't we see it?

To understand why the calculation does not work
simply on $\IR\uu 4$, let us recall the basic
framework for understanding quantum corrections
to large-$\JJM$ observables in effective field theory, as done in \cite{Hellerman:2017veg,Hellerman:2015nra,Monin:2016jmo}.
As in \cite{Hellerman:2017veg,Hellerman:2015nra,Monin:2016jmo} one can regularize and renormalize the
effective action at
an energy scale $\L$ parametrically below the UV
scale $E\lrm{UV}$, while keeping $\L$ larger
than the infrared scale.  The theory then has
a $1/ \JJM$ expansion as long as
\begin{align}
\begin{split}
E\lrm{UV} 
\gg
 E\lrm{IR}\ ,
\end{split}\label{CriterionvA}
\end{align}
which is satisfied so long as
\begin{align}
\begin{split}
E\lrm{UV} = \JJM^p\cc E\lrm{IR},\qquad p>0\ .
\end{split}\label{CriterionvB}
\end{align}
In the present theory, we have
\bbb
E\lrm{UV} = \langle |\phi| \rangle \ .
\eee
On $\IR\uu 4$, the only infrared scale is 
\begin{align}
\begin{split}
E\lrm{IR} 
 = \left|x_1-x_2\right|\uu{-1},
\end{split}
\end{align}
so our criterion \rr{CriterionvA} becomes
\begin{align}
\begin{split}\label{CriterionvC}
\langle |\phi|\rangle \gg  \left|x_1-x_2\right|\uu{-1}
\ .
\end{split}
\end{align}

By taking $\JJM$ large, we can indeed make
$\langle |\phi| \rangle$ as large as desired in
the region containing the points $x\ll 1$
and $x\ll 2$, as we can see from the classical solution
\rr{FinalSolnParametrization}.  
However we must be cautious: The VEV
of $|\phi|$ is a local, rather than a global
quantity, and in the classical solution \rr{FinalSolnParametrization}, the VEV $\left< |\phi| \right>$
falls to zero far enough away from the sources.
The effective theory cannot be used straightforwardly,
because the space $\IR\uu 4$ is never entirely in the regime
of validity of the effective theory by the criterion \rr{CriterionvC}.

But this is not fatal: The criterion \rr{CriterionvC} is
only a sufficient criterion, not a necessary one.  In a
conformal theory, there is clearly a looser criterion
that is still sufficient to render the amplitude under control
at large $\JJM$.  We can control large-$\JJM$ corrections
so long as the criterion \rr{CriterionvC} holds in
any conformal frame at all, not necessarily the $\IR\uu 4$ conformal frame.
In particular, if we conformally
transform to a sphere of radius $r \gtrsim |x\ll 1 - x\ll 2|$, in which the
points $x\ll{1}$ and $x_2$ have an $O(1)$ angular separation, 
then criterion \rr{CriterionvC} holds in the conformally
transformed frame, which includes a Weyl transformation
of the field $\phi$:
\begin{align}
\begin{split}
\phi\ll{\Sfour} = \left[\operatorname{\mathop{det}}\left ( {{\pp x\pr}\over{\pp
x}} \right )\right]\uu{-1}\cc \phi\ll{\IR\uu 4}\ ,
\label{WeylTransformExpr}
\end{split}
\end{align}
where $x\pr$ are the coordinates on the sphere and
$x$ are the coordinates on $\IR\uu 4$.

If we were to calculate ${\cal B}\ll n$
up to and including terms of order $\JJM\uu 0$, we would need 
to perform the conformal transformation explicitly,
in order to compute the fluctuation determinant of fluctuations around the classical solution, and to compute the
integral of the Wess--Zumino term.  Since we only want to compute up to order $\log(\JJM)$, the
situation is simpler.  In the conformal frame
of $\IR\uu 4$, the Wess--Zumino term is singular
at infinity; in the conformal
frame of the sphere, the contributions to the
$\log(\JJM)$ term from the Wess--Zumino term
and determinant are manifestly finite with higher corrections under control by
${1/ \JJM}$ suppression.  It is clear, now, what must have happened to the $\log(\JJM)$ term in flat $\IR\uu 4$: It is
hiding in the determinant, in the region where the effective theory has broken down.  But we can recover it by
conformally transforming to $\Sfour$.

Note that the superconformal $S\uu 4$ we will be using
is the maximally supersymmetric one preserving the
full $SU(2) \times U(1)$ R-symmetry and
the entire conformal $SO(5,1)$ isometry group, rather than
the smaller group preserved by the 
$D$-term deformation used to compute the vacuum
$S\uu 4$ partition function by localization in  \cite{Pestun:2007rz} and used to compute correlators in \cite{Baggio:2014ioa,Baggio:2014sna,Baggio:2015vxa,Gerchkovitz2017,Rodriguez-Gomez:2016ijh,Baggio:2016skg,Pini:2017ouj,Baggio:2017aww,Billo:2017glv}.  While we make use
of the results from these methods later in the paper
to check our large-$n$ predictions, we will never deform the supergravity
background from the maximally symmetric one.  Therefore there is no curvature-dependent
contact-term ambiguity in the structure of our chiral ring;
our $S\uu 4$ background is simply equivalent to
$\IR\uu 4$ by a change of variables, not by a
nontrivial $D$-term deformation.  The only change in 
the effective Lagrangian induced by the Weyl
transformation of the  background,
apart from the curvature of the sphere,
is the direct curvature coupling $\Delta {\cal L} = {1\over 6}
{\tt Ric}\ll 4 \cc |\phi\lrm{unit}|\sqd$ for a scalar field.

Rather than working out the solution $\phi\ll{S\uu 4}$ directly on the sphere,
we simply refer to the formula \rr{WeylTransformExpr}
for the conformally transformed
solution on $\IR\uu 4$.  We would need the detailed form of the expression in 
the $S\uu 4$ frame in
order to compute the order $n\uu 0$ terms in ${\cal B}\ll n$,
but since we are only computing to order $\log(\JJM)$ in
the present article, we shall need to understand only
certain qualitative features of the solution, and the expression 
\rr{WeylTransformExpr} suffices.

\section{Anomaly terms}\label{AnomalySec}

In section \ref{HigherDerivConformal},
we found that no superconformal
effective term can contribute larger than $\JJM\uu 0$.
However the Wess--Zumino terms cannot be written as superconformal
terms in the effective action, and they evade
the analysis in section \ref{HigherDerivConformal}.  There are Wess--Zumino terms for the Weyl symmetry and $U(1)$ $R$-charge, which
are needed to compensate the difference between the 
anomaly coefficients of the underlying CFT and the 
effective theory of the Coulomb branch.  They cannot be written as superconformal
terms (or conformal terms at all) in superspace, because
they explicitly break the Weyl symmetry and R-symmetry
of the action, in order to compensate the variation of the
path-integral measure of those same symmetries (though see \cite{Dine:1997nq} for an
explicitly supersymmetrization of the anomaly terms in flat space.)

In order to compute to $O(\log(\JJM))$ accurately, then, we must write the Wess--Zumino terms with some care paid
to their normalization.  We will focus solely on the normalization of the coefficient of the $O(\log(\JJM))$
contribution to the quantum effective action.  There is also an order $J\uu 0$ term, given by a
nontrivial integral.  Since we will only calculate up to and including order $\log(\JJM)$ in the effective action in the
present article, this integral will be unnecessary.  We will see that the form of the order 
$\log\JJM$ term 
 is simple and comes only from the background curvature in the form of the Euler density.

\subsection{Form of the Wess--Zumino terms}\label{WZTermForm}

Let us start with the Wess--Zumino effective action
which captures the conformal and $U(1)_R$-symmetry anomalies
in general $\mathcal N=1$ theories
 \cite{Komargodski:2011vj,Schwimmer:2010za,Luty:2012ww,Bobev:2013vta}.
 The explicit form of the action 
in Lorentzian signature is given by 
\cite{Bobev:2013vta}
 \begin{align}
S\lrm{WZ}
= \int \sqrt{g} \cc d\uu 4 x \cc {\cal L}\lrm{WZ}\ ,
\end{align}
with
\begin{align}
\begin{split}\label{LWZ}
{\cal L}\lrm{WZ} \equiv{}&\dilaton\left(
\Delta c^{\text{[KS]}}  W^2(g) -\Delta a^{\text{[KS]}}  E_4(g)-6\Delta c^{\text{[KS]}}  F^2\right)
\\&
+\beta\left[2\left(5\Delta a^{\text{[KS]}}-3\Delta c^{\text{[KS]}}\right) F\tilde F 
+\left(\Delta c^{\text{[KS]}}-\Delta a^{\text{[KS]}}\right)
R\tilde R
\right]
\\&-\Delta a^{\text{[KS]}}\left[
4\left(R^{\mu\nu}(g)-\frac12 R(g)g^{\mu\nu}\right)
\partial_\mu\dilaton \partial_\nu\dilaton
-2\left( \partial \dilaton\right)^2\left(2\Box \dilaton-\left(\partial\dilaton\right)^2\right)
\right]
,
\end{split}
\end{align}
where in our context $\Delta c\equiv c_{\rm CFT}-c_{\rm EFT}$ and
 $\Delta a\equiv a_{\rm CFT}-a_{\rm EFT}$.
Note that the normalization of the $a$- and $c$-coefficients used here and in  \cite{Komargodski:2011vj,Schwimmer:2010za,Luty:2012ww,Bobev:2013vta}, which  we denote by the superscript [KS],
differs from the one used in
\eqref{alphaAEFJ}
and the reference \cite{Anselmi:1997ys}.
The realtionship between the two (see section \ref{AnomalyConventionsAndValues} of the Appendix) is a factor of $16\pi^2$, {\it i.e.}, 
\begin{align}
\begin{split}\label{NormalizationKSAEFJ}
\left(a,c\right)^{\text{[KS]}} = \frac{1}{16\pi^2} \left(a,c\right)^{\text{[AEFJ]}}.
\end{split}
\end{align}
In the expression \rr{LWZ}, we have only written the bosonic component of these terms, rather than a fully supersymmetric
contribution to the Lagrangian.  It has been argued \cite{Dine:1997nq,Schwimmer:2010za} that such terms can be written as manifestly supersymmetric
but Weyl- and R-symmetry-violating terms in superspace.  The fermionic contributions are suppressed by
powers of $M\lrm{UV} \propto |\phi|\sqd \propto \JJM$, and so do not contribute even at order $\JJM\uu 0$.

\subsection{Evaluating the Wess--Zumino term on $\Sfour$}

So now let us evaluate the contribution of the Wess--Zumino term on $\Sfour$ directly.  As we explained above, this rather than the calculation on $\IR\uu 4$, is the perturbatively controlled calculation.

\heading{Euler coupling of the modulus on $\Sfour$}

On $\Sfour$  it is clear that there is only one contribution to the $\log(\JJM)$ term, and it is topological, just being proportional to a constant times the Euler density.   In order to compute the correct normalization
of the contribution of the
Euler term to the anomaly, we will use a few 
facts about the geometry of the four-sphere,
which we have written in the Appendix \ref{SecFourSphere}.

The natural normalization of the Euler density would be the "integer normalization" $E\ll 4\uu{\IZ}$, in which the
integral of the Euler density is simply the Euler number $\chi$ of the space:
\begin{align}
\begin{split}
\int
d^4x \cc \sqrt{|g|}\cc E\ll 4\uu\IZ = \chi\in\IZ\ ,
\end{split}
\end{align}
which equals $+2$ for the sphere $S\uu 4$, so the numerical value of $E\ll 4\uu\IZ$ for a sphere must be
\begin{align}
\begin{split}
E\ll 4\uu\IZ = {2\over{\operatorname{\mathop{Area}}({S\uu 4})}} = {3\over{4 \pi\sqd r\uu 4}}\ ,
\end{split}
\end{align}
where we have used the formula for the area of a four-sphere of radius $r$, ${\operatorname{\mathop{Area}}({S\uu 4})} = {{8\pi\sqd r\uu 4}/ 3}$. 
   For better or worse, the integer-normalization convention for the Euler density is not much used.  In \cite{Komargodski:2011vj}, the normalization of the Euler density is defined as
\begin{align}
\begin{split}\label{KomargodskiSchwimmerEulerDensityNormalization}
E\ll 4\uu{[\rm KS]} \equiv  R_{\mu\nu\rho\sigma}R^{\mu\nu\rho\sigma} -4 R_{\mu\nu}R^{\mu\nu} +R^2,
\end{split}
\end{align}
of which the numerical value for the four-sphere is
\begin{align}
\begin{split}
E\ll 4\uu{[\rm KS]} \equiv {{24}\over {r\uu 4}} = 32\pi\sqd\cc E\ll 4\uu\IZ\ .
\end{split}
\end{align}
so the relation between the two normalizations is
\begin{align}
\begin{split}
E\ll 4\uu{[\rm KS]} = 32\pi\sqd\cc E\ll 4\uu\IZ \ , \qquad E\ll 4\uu\IZ = {1\over{32\pi\sqd}}\cc E\ll 4\uu{[\rm KS]}\ .
\end{split}
\end{align}
Rewriting the   Wess-Zumino coupling\footnote{We have also
changed the sign of the term, which in  \cite{Komargodski:2011vj} was written as a term in a Lorentzian action, as appropriate
to the context of dilaton scattering studied in that paper.  For the purposes of a path integral on $S\uu 4$, the relevant
action is the Euclidean one, which is the negative of the Lorentzian action after Wick-rotation.}
of  \cite{Komargodski:2011vj}
in terms of the somewhat more intuitive $E\ll 4\uu\IZ $, the Euler coupling term of the dilaton is, in Euclidean signature,
\begin{align}
\begin{split}
\label{EulerCouplingExpression}
{\cal L}_{\text{WZ}}^{\text{Euler coupling}} = 
+1\cc \Delta a\uu{[\rm KS]}\cc E\ll 4\uu{[\rm KS]} \cc\dilaton
=  32\pi\sqd\cc \Delta a\uu{[\rm KS]}\cc E\ll 4\uu\IZ\cc\dilaton
=2\cc \Delta a\uu{[\rm AEFJ]}\cc E\ll 4\uu\IZ\cc\dilaton
.
\end{split}
\end{align}
Here $\Delta a$ is the difference between the $a$-anomaly coefficient of the full interacting CFT, and that of the 
infrared-free effective theory on the moduli space of supersymmetric vacua, $\Delta a \equiv a\lrm{CFT} - a\lrm{EFT}$.  By virtue
of the Komargodski--Schwimmer $a$-theorem, this number $\Delta a$ is always positive.
The dilaton $\dilaton$ is normalized in \cite{Komargodski:2011vj} such that $\eexp(- \dilaton)$ transforms as a scalar of dimension 
$+1$; in a supersymmetric moduli space in four dimensions, then, the field $\eexp(-\dilaton)$ is the modulus $|\phi| = |\phi\lrm{unit}|$ that
spontaneously breaks the scale invariance, giving us the identification
\begin{align}
\begin{split}
\dilaton = - \log\frac{\left|\phi\right|}{\mu},
\end{split}\label{DilatonFromModulusPhi}
\end{align}
where $\m$ is an arbitrary mass scale.

\heading{Boundedness of the $O(\JJM\uu 0)$ term}

In order to compute the full contribution of the Wess-Zumino coupling including the $\JJM\uu 0$ term, we would need to compute the entire
profile of $\dilaton$, given by substituting the classical solution  \rr{FinalSolnParametrization}
Since we are not attempting to compute the order $n\uu 0$ term, we will not need to do the integral at all: 
The classical solution \rr{FinalSolnParametrization} for $|\phi|$ has a fixed scaling limit as $\JJM\to\infty$, and so its logarithm can be decomposed
as the sum of an $x$-independent piece, and a piece bounded by order $|\JJM|\uu 0$:
\begin{align}
\begin{split}\label{PhiMagDecomp}
|\phi| = \JJM\uu\hh\cc 
|
\widehat\phi
|
\ ,
\end{split}
\end{align}
where $|
\widehat\phi
|$ is of order $\JJM\uu 0$ in the $S\uu 4$ conformal frame, away from the singularity at the insertion points.

The singularities at the insertion points $x$ and $y$ at first sight seem like they might cause the large-$\JJM$ expansion to break down, but they do not.  In general, ultraviolet singularities should never be a problem, as we regularize and renormalize our theory at a distance scale
$\L\uu{-1} \gg |\phi|\uu{-1}$.  In this case, that is not even necessary: Due to cancellations of the most naively
singular terms and the fact that the solution is complex rather than real, the integrated Wess--Zumino action does
not diverge at the insertion points, and the integral is finite as the regulator is removed.

Thus using \rr{PhiMagDecomp} we find that the Wess--Zumino Lagrangian density can be written as
\bbb
{\cal L}\lrm{WZ} = {\cal L}_{\text{WZ}}^{\text{Euler coupling}} + O(\JJM\uu 0)\ ,
\eee
where the $O(\JJM\uu 0)$ piece is finite, and we can discard it at our desired order of precision.
Then, combining \rr{DilatonFromModulusPhi} with the decomposition \rr{PhiMagDecomp} gives
\begin{align}
\begin{split}
\dilaton = - \hh \log(\JJM) + O(\JJM\uu 0)\ ,
\end{split}\label{DilatonFromModulusPhiDECOMP}
\end{align}
and further using  \rr{EulerCouplingExpression} and
the fact that the Euler number of the sphere is $\chi\ll{\Sfour} = +2$, we have
\begin{align}
\begin{split}
S\lrm{WZ} = - \a \log(\JJM) + O(\JJM\uu 0)\ , 
\end{split}\label{WessZuminoActionToLogJ}
\end{align}
where
\begin{align}
\begin{split}
\a = 2\cc (a\lrm{CFT} - a\lrm{EFT})\uu{\rm [AEFJ]}\ .
\end{split}\label{AlphaCoefficientValueFirstTime}
\end{align}

So then the Euclidean path integral gets a multiplicative contribution of $Z\lrm{ {{classical}\atop{WZ}}}$,
\begin{align}
\begin{gathered}
Z\ll n= Z\lrm{{{leading}\atop{factors}}}  \cdot 
\big [O(n\uu 0)+ O(n\uu {-1}) \big ] ,
\\
Z\lrm{{{leading}\atop{factors}}} \equiv
Z^{\text{free-field}} Z\lrm{ {{classical}\atop{WZ}}},
\qquad
 \eexp\left({- S\lrm{{{classical}\atop{Euclidean~WZ}}}} \right)= \JJM^{\a}\cdot 
 O(n\uu 0),
\end{gathered}
\end{align}
 and so
\begin{align}
\begin{split}
 Z\ll n&= \JJM\uu\a\cc Z^{\text{free-field}}
 \left [O(n\uu 0)+ O(n\uu {-1}) \right ] 
 = \left(  {{|{\bf N}|\ll\co}\over{2\pi |x - y|}}  \right)
 \uu{2\JJM}(\JJM)!  \JJM\uu\a  \left [O(n\uu 0)+ O(n\uu {-1}) \right ],
\end{split}
\end{align}
 where $Z\ll n$ is the full CFT path integral with sources \rr{PathIntWithSources},
and $Z^{\text{free-field}}\ll n$ is the corresponding free-field approximation to the path integral for the two-point functions in the effective theory.

\section{Localization in rank-1 theories with marginal couplings}\label{localization}

Following   \cite{Baggio:2014ioa,Baggio:2014sna,Baggio:2015vxa,Gerchkovitz2017}  we briefly review how to  compute by supersymmetric localization two-point functions of various four-dimensional $\mathcal N\geq2$ superconformal field theories in $\mathbb R^4$.  We will
then apply the results of  \cite{Baggio:2014ioa,Baggio:2014sna,Baggio:2015vxa,Gerchkovitz2017} to
Lagrangian conformal theories with gauge group $SU(2)$
(or $SO(3)$), and compare with our asymptotic expansion
of $\ampname\ll n$.  The two interacting conformal theories 
with marginal couplings are ${\cal N} = 4$ super-Yang--Mills with gauge group $SU(2)$ (or $SO(3)$), and
superconformal QCD with four hypermultiplets
in the fundamental representation ${\bf 2}$ of $SU(2)$,
$N\ll f = 4$.

\subsection{Relation of conventions}

For rank-one theories with $\D\ll\co = 2$, our $\ampname\ll n$ depends on the marginal parameter $\t,\tb$ and
is identified with the two-point function $G\ll{2n}(\t,\tb)$, up 
to powers of a normalization factor we denote by $\bf K$
such that
\begin{align}
\begin{split}
\co\uu{\rm here} 
= {\bf K}\uu{\D\ll\co}\cdot \co\ll 2 ^{\text{ref  \cite{Baggio:2014ioa,Baggio:2014sna,Baggio:2015vxa,Gerchkovitz2017}}}  \ ,
\end{split}
\end{align}
The dimension of the generator $\co$ is $\D\ll\co = 2$ for the two
 theories we consider in this section.  With the relative normalizations defined this way, the relationship of the two-point 
functions is
\begin{align}
\begin{split}
\ampname\ll n = \ampname\ll n(\t,\tb) = \eexp\left({{\cal B}\ll n(\t,\tb)}\right) =\left |{\bf K}\right|\uu{4n}
G\ll{2n}(\t,\tb).
\end{split}
\end{align}
With this identification we will review the computation
of correlation functions in  \cite{Baggio:2014ioa,Baggio:2014sna,Baggio:2015vxa,Gerchkovitz2017} and
then compare with our own results when $n$ is large.

\subsection{Method of \cite{Baggio:2014ioa,Baggio:2014sna,Baggio:2015vxa,Gerchkovitz2017}}

To calculate two-point functions on $\IR\uu 4$, one first needs the ${\Sfour}$ partition function $Z_{\Sfour}(\tau,\bar\tau)$ associated with
the $\mathcal N\geq2$ SCFT action $S_{\text{SCFT}}$ deformed by the chiral ring generators ${\mathcal O}_i$,
	\begin{align}
		\begin{split}
S_{\text{SCFT}} \to S_{\text{SCFT}} - \frac{1}{32\pi^2}\left(\int d^4x \cc\cc d^4\theta \cc \cc
\mathcal E \cc\cc \sum_i\tau_i {\mathcal O}_i + \mathrm{c.c.}\right),
		\end{split}
	\end{align}
where $\mathcal E$ is the  chiral density of $\mathcal N=2$ supergravity and $\tau_i$ are holomorphic coupling constants.
Since this deformed theory preserves $\mathfrak{osp}(2|4)$, the massive $\mathcal N=2$ supersymmetry algebra on ${\Sfour}$,
the associated partition function  can be computed exactly by localization.

In the case of theories with one-dimensional Coulomb branch,\footnote{For theories with multi-dimensional Coulomb branch, see  \cite{Baggio:2015vxa,Gerchkovitz2017}.} the two-point functions of the chiral ring operators defined by 
\begin{align}
\begin{split}\label{G2n}
G_{2n}\left( \tau,\bar\tau \right) \coloneqq \left< {\mathcal O}^n(  0) \bar {\mathcal O}^n(  \infty)\right>,
\qquad 
\bar {\mathcal O}^n(  \infty)\coloneqq \lim_{\left| x\right|\to\infty}\left| x\right|^{2n\D\ll\co}  \bar{\mathcal O}^n(x),
\end{split}
\end{align}
can be computed systematically in the following way. 
First, one computes  derivatives of  $Z_{\Sfour}(\tau,\bar\tau)$ and constructs a matrix $M$ whose $(m,n)$-entry $(m,n=0,1,2,\cdots)$ is given by
	\begin{align}\label{Eq111}
		\begin{split}
M_{m,n} \coloneqq \frac{1}{ Z_{\Sfour}(\tau,\bar\tau)} \partial_\tau ^m \partial_{\bar\tau}^n Z_{\Sfour}(\tau,\bar\tau).
		\end{split}
	\end{align}
Then, the two-point functions \eqref{G2n} can be obtained as\footnote{Equation \eqref{Eq222} satisfies the $tt^*$ equation \cite{Papadodimas:2009eu,Baggio:2014ioa}.}
	\begin{align}
		\begin{split}\label{Eq222}
G_{2n}(\tau,\bar\tau) = 16^n \frac{\mathop{\mathrm{det}} M_{(n)}}{\mathop{\mathrm{det}} M_{(n-1)}},\qquad n=1,2,3,\cdots,
		\end{split}
	\end{align}
where $M_{(n)}$ is the upper-left $(n+1)\times(n+1)$ %
submatrix of $M$.

\subsection{The case of $SU(2)$ $\mathcal N=4$ SYM}
$\mathcal N=4$ super-Yang--Mills with gauge group $SU(2)$  has a very simple partition function \cite{Pestun:2007rz},
\begin{align}
	\begin{split}\label{symz}
Z^{\mathcal N=4}_{\Sfour}(\tau,\bar\tau ) = \frac{1}{4\pi\left( \Im \tau  \right)^{\nicefrac32}},
	\end{split}
\end{align}
where in this case $\tau$ is the complexified Yang--Mills coupling\footnote{Again, we note we use
this convention regardless of matter content, our sole deviation from the conventions of
\cite{Seiberg:1994aj, Seiberg:1994rs}, who define $\tau\ll{{\cal N} = 2~{\rm SQCD}} = \frac{\theta_{\text{YM}}}{\pi} + \frac{8\pi i}{g^2\lrm{YM}}$.}
	\begin{align}
		\begin{split}
\tau = \frac{\theta_{\text{YM}}}{2\pi} + \frac{4\pi i}{g^2}.
		\end{split}
	\label{TauInTermsOfGYMAndTheta}
	\end{align}
From \eqref{symz} one deduces
\begin{align}
	\begin{split}\label{ExactTwoPtFuncNEqualsFour}
G_{2n}(\tau,\bar\tau) = \frac{(2n+1)!}{\left(\Im \tau\right)^{2n}},
\qquad
\ampname\ll n = |{\bf K}|\uu {4n} \cc G_{2n}=\left(\cc
{{|{\bf K}|\sqd}\over{\Im \tau }} \right)\uu{2n}\cc (2n+1)!\ .
	\end{split}
\end{align}
%
At large $n$, the logarithm of $G_{2n}(\tau,\bar\tau) $ becomes
\begin{align}
	\begin{split}
\log \left[G_{2n}(\tau,\bar\tau) \right]
= 2n\log n+n\left(2\log 2-2-2\log\left(\Im\tau\right)\right) +\frac32\log n
+O(1).
	\end{split}
\end{align}
This matches our prediction \rr{MainFormulaForAmpname} up to the order
to which we were retaining terms, order $n\uu 0$.

\subsection{Numerical analysis of SQCD with $N\ll f = 2 N\ll c = 4$}

One would like to find as many other rank-one superconformal ${\cal N} = 2$
theories as possible for which we could compare our general
results with two-point functions computed via
localization.  Unfortunately, there are not many
examples in the literature that have
been worked out already.  In \cite{Baggio:2014sna,Baggio:2014ioa,Gerchkovitz2017}
the authors study the example of $SU(2)$ ${\cal N} = 2$ SQCD
with four doublet hypermultiplets.  Even for that
relatively simple theory the sphere partition function and $G\ll{2n}$ for low values of $n$, have a complex $\t$ dependence with a 
 nonperturbative definition via
an integral, but not one that is simple to write in closed form. 
 It is possible however, to evaluate
the two-point function $\ampname\ll n$ numerically for any value of $\t$, to good enough
accuracy to extract the coefficients of the large-$n$ expansion of $\log(\ampname\ll n)$
with some precision.  In particular, we are able to extract the coefficient $\a$ of $\log(n)$
and compare it to the prediction of the EFT analysis.

The sphere partition function 
of $SU(2)$ $\mathcal N=2$ SQCD with four fundamental hypermultiplets
is given by \cite{Nekrasov:2002qd,Alday:2009aq}
\begin{align}
	\begin{split}
Z^{\mathcal N=2}_{\Sfour} (\tau,\bar\tau) = 
\int^\infty_{-\infty} da
a^2
e^{-4a^2\Im \tau}
\frac{\abs{G(1+2ia)}^4}{\abs{G(1+ia)}^{16}}
\abs{Z_{\text{inst}}(ia,\tau)}^2,
	\end{split}
\end{align}
where the function
$G(x)$ is the Barnes $G$-function \cite{barnes},
and $Z_{\text{inst}}(a,\tau)$ is the instanton partition
function, which is expanded as\footnote{See {\it e.g.} \cite{Aniceto:2014hoa} for higher order terms in this expansion.}
\begin{align}
\begin{split}
Z_{\text{inst}}(ia,\tau)
=
1+\frac12\left(a^2-3  \right)e^{2\pi i\tau} + O\left(e^{4\pi i \tau}\right).
\end{split}
\end{align}
For 
the sake of simplicity we concentrate on the region
$\Im\tau \geq1$ and 
ignore all the instanton corrections.
The zero-instanton sector of the sphere partition function does not
depend on $\operatorname{\mathop{Re}} \tau$.
Using \eqref{Eq111} and \eqref{Eq222}, 
we  evaluate the two-point functions $G_{2n}$ 
up to an arbitrary order in $n$ for any value
 of $\Im \tau$.
In figure \ref{fig:aa} we have plotted a particular
combination of logarithms of $G\ll{2n}$ that 
comprise the left-hand side of the sum rule
\rr{SQCDSumRuleFOrd2},
approximating the $S\uu 4$ partition function with 
the perturbative part alone.
The asymptotic value should be $-4$ for any value
of $\t$, if we start
the recursion relations with the full $S\uu 4$ partition
function with instanton corrections included.  That
is, in the fully instanton-corrected theory we should
have \rr{SQCDSumRuleFForDeltaEquals2Ord2}.

\subsection{Comparison of exact results with the
large-$J$ expansion}

Now we will compare results, using the value of the $\a$-coefficient computed in the Appendix.  In eq. \rr{NEqualsFourAlpheCoefficient}.
we computed the $\a$-coefficient for ${\cal N} = 4$ super-Yang-mills
with gauge group $SU(2)$, and we found
\bbb
\a\ll{{\cal N} = 4~{\rm SYM},~G=SU(2)} = 1\ .
\een{NEqualsFourAlpheCoefficientPRECAP}

We therefore expect 
\begin{align}
\begin{split}
\ampname\ll n = n\uu{+1}\cc (2n)!  \left ( {{|{\bf N}\ll \co |}\over{2\pi}} \right )\uu{2n} \cc \left [ O(n\uu 0) + O(n\uu {-1}) + \cdots \right   ]  \ .
\end{split}
\end{align}
The exact formula \rr{ExactTwoPtFuncNEqualsFour}
can be written as
\begin{align}
\begin{split}
 G\ll{2n} = n \uu{+1}\cc (2n)!  \left ( {{|\widehat{{\bf N}}\ll \co |}\over{2\pi}} \right )\uu{2n}  \left (  2 + {1\over n}  \right ) \ ,
\end{split}
\end{align}
which agrees with the form of our asymptotic expansion, with
\begin{align}
\begin{split}
\D\ll\co = 2\ , 
\qquad
 \a = +1\ , \qquad \tilde{\ampname}\ll n = 2 + {1\over n}\ ,
 \qquad \widehat{{\bf N}}\ll \co  \equiv 
{\bf K}\uu{-1} {\bf N}\ll\co\ .
\end{split}
\end{align}

\heading{The case of ${\cal N} = 2$ SQCD with $N\ll f = 2N\ll c = 4$}

For conformal SQCD with $N\ll f = 2N\ll c = 4$,  we have $\D\ll \co = 2$ and in equation \rr{SQCDAlphaCoefficientValue} of the Appendix we have calculated 
\begin{align}
\begin{split}\label{SQCDAlphaCoefficientValuePRECAP}
\a\ll{{\rm SCQCD},~G=SU(2)} = \frac32\ .
\end{split}
\end{align}
In this case, our data is only numerical, derived
from recursion relations starting from the perturbative approximation to the $S\uu 4$ partition function.
Therefore it is easier to check the
accuracy of the sum/product rules of sec. \ref{AsympSumRules}
than to fit the data to a curve.
We  expect the two-point functions to obey the sum
and product rules \rr{SumRuleF} and \rr{SQCDSumRuleG} with $\D\ll\co = 2,\cc\cc \a = \nicefrac32$,
\begin{align}
\begin{split}\label{SQCDSumRuleFForDeltaEquals2}
{\cal B}\ll {n + 2} + {\cal B}\ll n - 2 \cc {\cal B}\ll{n+1}
= {2\over n} -4\cc 
 n\uu{-2}
+ O(n\uu {-3})\ ,
\end{split}
\end{align}
%
%
%
%
which imply the individual limits
\begin{align}
\lim_{n\to\infty}
\left( {\cal B}\ll {n + 2} + {\cal B}\ll n - 2 \cc {\cal B}\ll{n+1} \right)&= 0\ ,
\label{SQCDSumRuleFForDeltaEquals2Ord0}
\\
\lim_{n\to\infty}  n\cc\left ( \cc
{\cal B}\ll {n + 2} + {\cal B}\ll n - 2 \cc {\cal B}\ll{n+1}
\cc \right) &= 2\ ,
\label{SQCDSumRuleFForDeltaEquals2Ord1}
\\
\lim_{n\to\infty}  \left[\cc
n\sqd\cc\left( \cc
{\cal B}\ll {n + 2} + {\cal B}\ll n - 2 \cc {\cal B}\ll{n+1} \cc \right)
 - 2\cc n  \cc\right] &= - 4\ ,
 \label{SQCDSumRuleFForDeltaEquals2Ord2}
\end{align}

\begin{figure}[!h]
\center
\includegraphics[width=40 em]{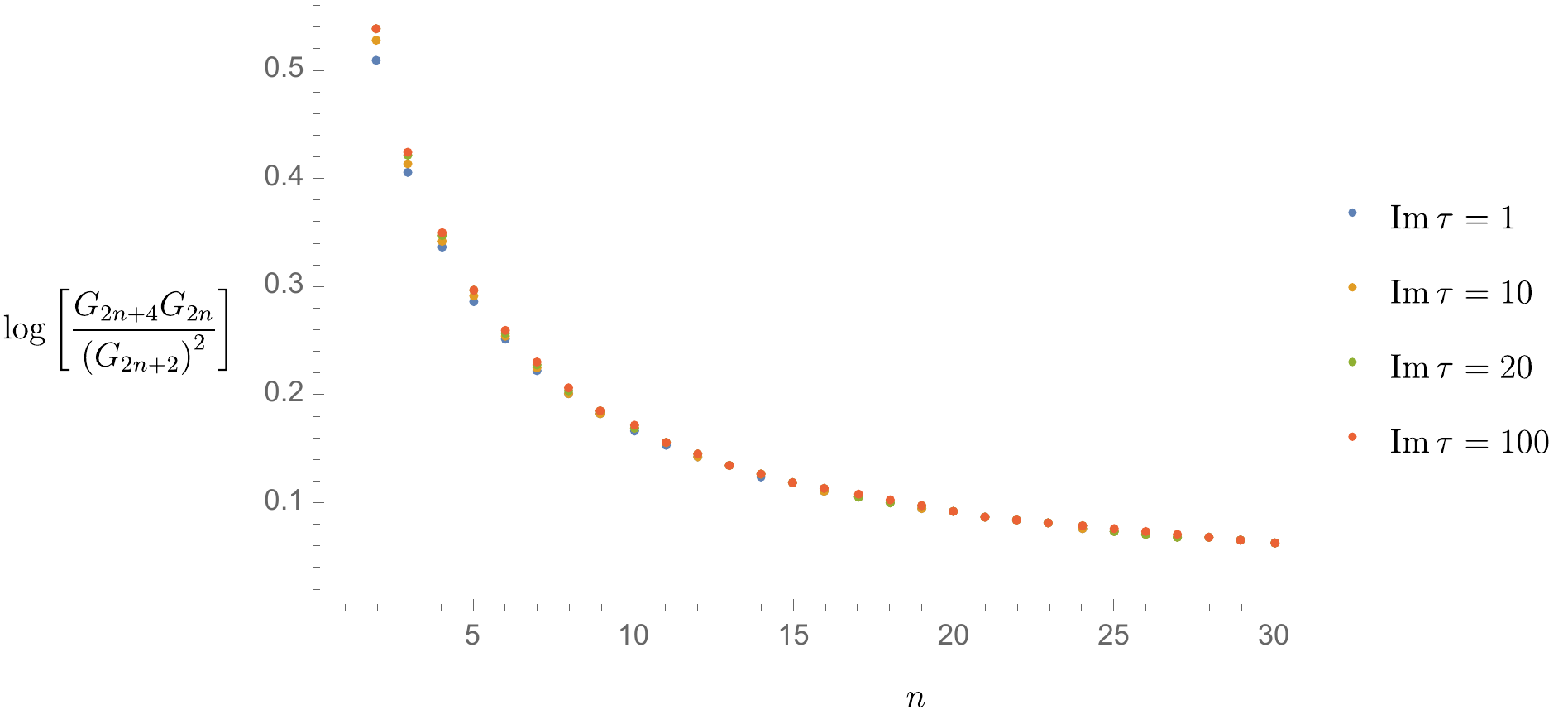}
\caption{Approximate values of 
the LHS of sum rule 
\rr{SQCDSumRuleFForDeltaEquals2Ord0}
in conformal SQCD with $G=SU(2)$ and $N\ll f = 4$, calculated via recursion relations 
from the $S\uu 4$ partition function, with instanton corrections omitted.}\label{fig:BB}
\end{figure}

\begin{figure}[!h]
\center
\includegraphics[width=40 em]{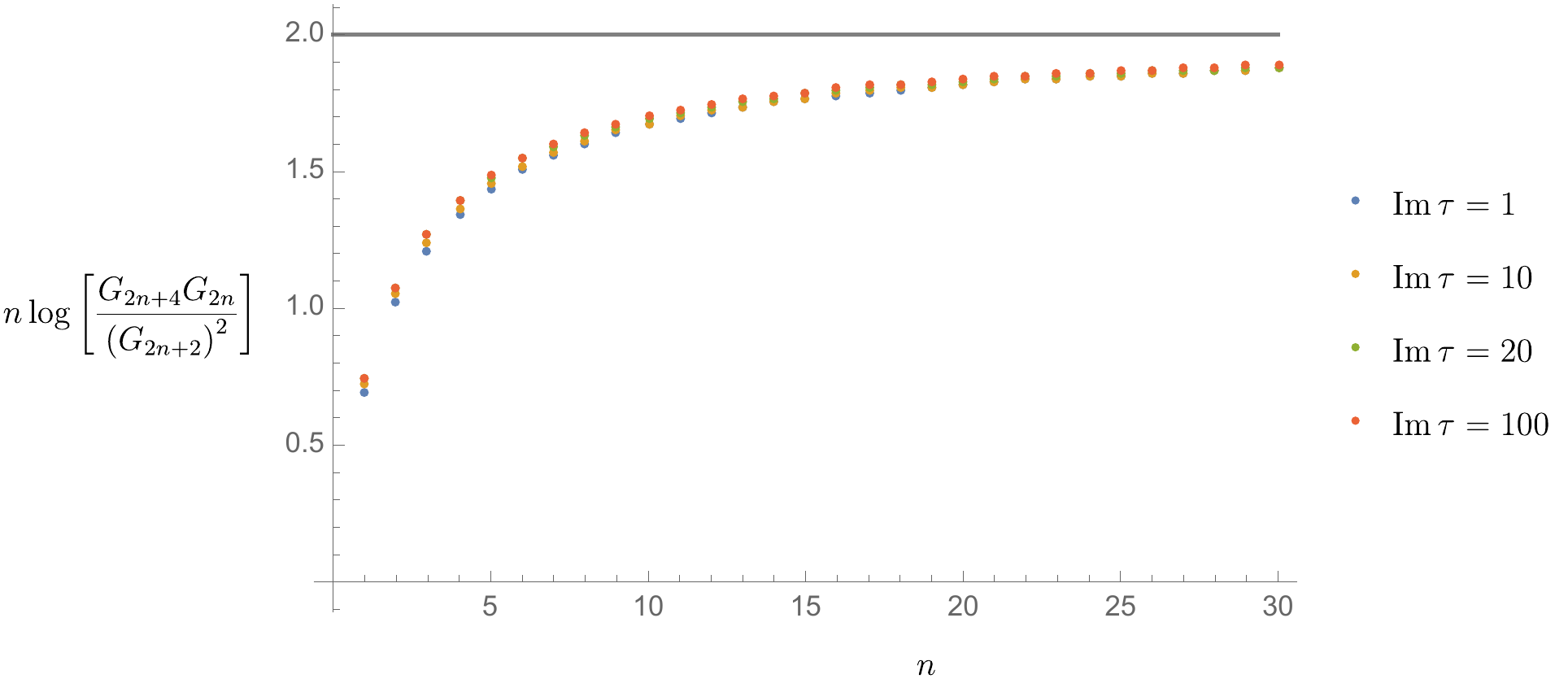}
\caption{Approximate values of 
the LHS of sum rule 
\rr{SQCDSumRuleFForDeltaEquals2Ord1}
in conformal SQCD with $G=SU(2)$ and $N\ll f = 4$, calculated via recursion relations 
from the $S\uu 4$ partition function, with instanton corrections omitted.}\label{fig:CC}
\end{figure}

\begin{figure}[!h]
\center
\includegraphics[width=40 em]{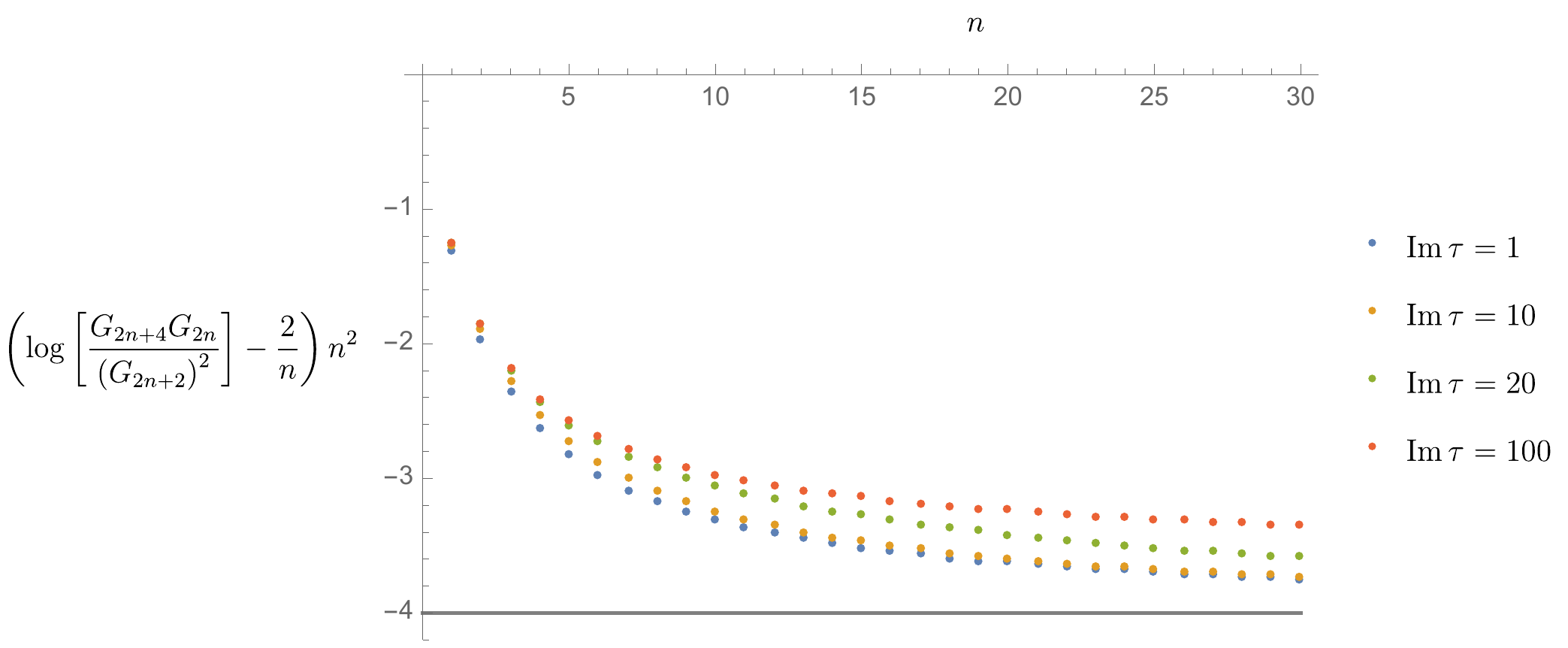}
\caption{Approximate values of 
the LHS of sum rule 
\rr{SQCDSumRuleFForDeltaEquals2Ord2}
in conformal SQCD with $G=SU(2)$ and $N\ll f = 4$, calculated via recursion relations 
from the $S\uu 4$ partition function, with instanton corrections omitted.    For the exact $S\uu 4$ partition function, with
all instanton corrections, our analysis predicts the LHS of \rr{SQCDSumRuleFForDeltaEquals2Ord2} should
approach $-4$ for any $\t$, as $n$ goes to infinity. It would appear unlikely that the asymptotic value of
of the sum rule is truly $-4$ for the no-instanton approximation to the $S\uu 4$ partition function, but at present the
authors have no theory of the error.}\label{fig:aa}
\end{figure}

In figures \ref{fig:BB}, \ref{fig:CC}, \ref{fig:aa} we plot the LHS of equations
\rr{SQCDSumRuleFForDeltaEquals2Ord0},
\rr{SQCDSumRuleFForDeltaEquals2Ord1},
\rr{SQCDSumRuleFForDeltaEquals2Ord2}, up
to $n = 30$, for various (purely imaginary) values of $\t$.
These values have been calculated by the recursion relations of  \cite{Baggio:2014ioa,Baggio:2014sna,Baggio:2015vxa,Gerchkovitz2017},
approximating the sphere partition function by its
perturbative piece alone.  Even in this
approximation, the large-$n$ prediction \rr{SQCDSumRuleFForDeltaEquals2Ord2} is close to $-4$
for $n$ of order $30$.  Note that the agreement is
best at $\t = i$, which is expected to have the
lowest threshold for the applicability of the large-$J$ approximation, as the gap above the massless sector is
highest  there. We do not know whether the omission of instanton corrections affects the true asymptotic
value of the LHS of the sum rule \rr{SQCDSumRuleFForDeltaEquals2Ord2}, or whether the sum rule would
indeed converge to $-4$ for sufficiently large $R$-charge, even without instanton corrections.

\newpage

\section{Conclusions}
\heading{Other theories with one-dimensional Coulomb branch}
There are many other theories with one dimensional
Coulomb branch (or more generally with
a single vector multiplet and massless hypers) without marginal coupling.  Since these do not have marginal
couplings, they are harder to do explicit calculations
with and we do not have results in the literature with
which we can easily compare.  In order to predict
correlation functions of $(\co)\uu n$ for large $n$,
we must know the dimension of the generator $\co$, 
the $a$-coefficient of the full CFT, and the
massless content of the effective theory on moduli space.


Rank-one SCFT have been the subject of
intensive recent study by \cite{Argyres:2015ffa,Argyres:2015gha,Argyres:2016xmc,Argyres:2016xua },
in which theories with one-dimensional Coulomb branch
were classified under broad conditions.

We make use of the beautiful results \cite{Argyres:2016xmc,Argyres:2016xua,Argyres:2015gha, Argyres:2015ffa} on the classification of rank-one superconformal field theories.  Actually we will do more than just "make use of" them: Table \ref{tabletable}
is created by copying directly\footnote{This table was created in part by copying the {\LaTeX} 
 code of table $1$ of \cite{Argyres:2016xmc}.  Table 
doing so with the intention of communicating our results for the $\a$-coefficients and their relation to \cite{Argyres:2016xmc}, 
in a context that is most easily understood
by the reader.  We do not claim as original work the creation of the content or appearance of our table \ref{tabletable} insofar as it overlaps
with table 1 of \cite{Argyres:2016xmc}.
According to our best understanding, this is a legitimate use of the work \cite{Argyres:2016xmc} under
the arXiv non-exclusive license to distribute, \url{https://arxiv.org/licenses/nonexclusive-distrib/1.0/license.html}.} a table
from \cite{Argyres:2016xmc}, but with our our own additional columns, giving data on the Wess--Zumino term
and the value of the $\a$-coefficient of the theory.

\begin{table}[ht]
\centering \small
$\def\arraystretch{1.0}
\begin{array}{clc|c|c|cc||cc|c}
\hline &
 \multicolumn{2}{l|}{\text{Coulomb branch:}} &
  \text{higgs} &
\multicolumn{1}{l|}{\text{massless}} &
\multicolumn{2}{l||}{\text{Central charges:}} &
\multicolumn{2}{l|}{\text{Wess-Zumino term:}} &
\multicolumn{1}{l}{\text{$\a$-coefficient:}}
\\[1mm]
&\text{$ {{\rm singul.}\atop{\rm type}}$} & \D\ll\co & \text{br. dim}
& {\rm hypers} & \ \ 24a\ \ &\ \ 12c\ \  &\ \  24a\lrm{EFT} \ \  & \ \ 24\Delta a \ \ & \ \ \a = 2\Delta a\ \
\\[1.5mm]
\hline\hline
&&&&&&&&&\\[-4.5mm]
&II^* & 6 
& 29 & 0 \GGO & 95 & 62 & 5 & 90 & 15/2 \\
&III^* & 4 
& 17 & 0\GGO  & 59 & 38 & 5 & 54 & 9/2 \\
&IV^* & 3 
& 11 & 0\GGO  & 41 & 26 & 5 & 36 & 3 \\
\rcy &I_0^* & 2 
& 5 & 0 \GGO & 23 & 14 & 5 & 18 & 3/2 \\
&IV & 3/2 
& 2 & 0\GGO & 14 & 8  & 5 & 9 & 3/4 \\
&III & 4/3 
& 1 & 0 \GGO  & 11 & 6 & 5 & 6 & 1/2 \\
&II & 6/5 
& 0 & 0 & \GGO 43/5 & 22/5  & 5 & 18/5 & 3/10 \\
\rcr \multirow{-9}{4mm}{\begin{sideways}$I_1$ series\qquad \  \end{sideways}}
&I_1 & 1 
& 0 & 0\GGO & 6 & 3 & 5 & 1 & 1/12 \\[.5mm]
\hline\hline
&&&&&&&&& \\[-4.5mm]
&II^* & 6 
& 16 & 5\GGO & 82 & 49 & 10 & 72 & 6 \\
&III^* & 4 
& 8 & 3\GGO  & 50 & 29 & 8 & 42 & 7/2 \\
&IV^* & 3 
& 4 & 2\GGO  & 34 & 19 & 7 & 27 & 9/4 \\
\rcy &\blue{I_0^*} & 2 
& 0 & 1 \GGO & 18 & 9 & 6 & 12 & 1 \\
\rcr \multirow{-6}{4mm}{\begin{sideways}$I_4$ series\qquad \ \end{sideways}}
&I_4 & 1 
& 0 & 0\GGO & 6 & 3 & 5 & 1 & 1/12 \\ [.5mm]
\hline\hline
&&&&&&&&& \\[-4.5mm]
&II^* & 6 
& 9 & 4 \GGO & 75 & 42 & 9 & 66 & 11/2 \\
&III^* & 4 
& ? & 2 \GGO  & 45 & 24 & 7 & 38 & 19/6 \\
&\red{IV^*} & 3 
& 0 & 1\GGO  & 30 & 15 & 6 & 24 & 2 \\
\rcr \multirow{-5}{4mm}{\begin{sideways}$I_1^*$ series\qquad\  \end{sideways}} 
&I_1^* & 2 
& 0 & 0\GGO& 17 & 8 & 5 & 12 & 1 \\[.5mm]
\hline\hline
&&&&&&&&& \\[-4.5mm]
&II^* & 6 
& ? & 3\GGO & 71 & 38 & 8 & 63 & 21/4 \\
&\red{III^*} & 4 
& 0 & 1\GGO & 42 & 21 & 6 & 36 & 3 \\
\multirow{-4}{4mm}{\begin{sideways}$\scriptstyle{IV^*_{\scriptscriptstyle Q=1}}$ \small{ser.}\quad \  \end{sideways}} 
&IV^*_{Q=1} & 3 
& 0 & 0\GGO & 55/2 & 25/2 & 5 & 45/2 & 9/4 \\[.5mm]
\hline\hline
\rcy &&&&&&&&& \\[-4.5mm]
\rcy &\blue{I_0^*} & 2 
& 0 & 1\GGO  & 18 & 9 & 6 & 12 & 1 \\
\rcr \multirow{-3}{4mm}{\begin{sideways}$I_2$ ser.\quad\ \  \end{sideways}} 
&I_2 & 1 
& 0 & 0\GGO & 6 & 3 & 5 & 1 & 1/12 \\[.5mm]
\hline\hline
\end{array}$
\caption{Argyres, Lotito, L\"u and Martone's partial list of rank-1 $\cN=2$ SCFTs.  This table has been copied directly (at the level
of the {\LaTeX} code even) from  \cite{Argyres:2016xmc}, to clarify the identification of theories, which are
labelled exactly as in that reference.  
We have 
added the three columns on the right, including the $\a$-coefficient.
The column `massless hypers' denotes
the number of hypermultiplets massless at a generic point on the Coulomb branch, a situation
referred to in \cite{Argyres:2016xmc} as an `enhanced Coulomb branch' (ECB) if the number is nonzero.
}
\label{tabletable}
\end{table}



\heading{Conclusions}

In this paper we have analyzed the large-quantum-number
expansion of two-point functions of operators $\co\ll\D\uu n$,
where $\co\ll\D$ is the holomorphic generator of a
Coulomb branch chiral ring in a rank-one superconformal
field theory.  To do this, we have followed earlier works
and used the effective field theory governing the
large-$\JJM$ sector of the Hilbert space.  As in
the previous paper \cite{Hellerman:2017veg} on the superconformal large-$\JJM$ expansion,
the relevant EFT is the effective dynamics of
the supersymmetric moduli space, which is governed
by spontaneously broken superconformal symmetry.  
We have used the Coulomb-branch EFT to
expand the two-point function
\begin{align*}
\begin{split}
\ampname\ll n \equiv  |x - y|\uu{ 2 \JJM}\cc\left< (\co (x) )\uu n
\cc  (\bar{\co}(y) )\uu n \right>
\end{split}
\end{align*}
 at large $R$-charge, \it i.e., \rm for $n\gg 1$.  The EFT predicts
 that $\ampname\ll n$ has
 an asymptotic expansion at large $n$, behaving as
 \begin{align*}
\begin{split}
\ampname\ll n = 
(\JJM)! \cc \left ({{|{\bf N}\ll\co|}\over{2\pi}} \right )\uu {2\JJM} \cc  \JJM\uu\a \cc  \tilde{\ampname}\ll n\ ,
\end{split}
\end{align*}
where  $\tilde{\ampname}\ll n$ approaches a constant as $n\to\infty$, and ${\bf N}\ll\co$ is an $n$-independent constant depending
on the normalization of the operator relative to the effective Abelian gauge coupling $g\lrm{eff}$.
We have calculated the exponent $\a$ and found that it is computed entirely
by the coupling between the Euler density of the sphere and the logarithm of the scalar modulus $|\phi|$.
This coupling is fixed by anomaly matching to be proportional to the difference between the 
$a$-anomaly coefficient of the underlying CFT and that of the EFT of massless moduli.  In
the conventions of \cite{Anselmi:1997ys}, this is
\begin{align*}
\begin{split}
\a = 2 \cc \left (a\lrm{CFT} - a\lrm{EFT}\right ) \uu{\rm[AEFJ]}\ .
\end{split}
\end{align*}
 In theories with a marginal coupling, we have used results from localization \cite{Baggio:2014ioa,Baggio:2014sna,Baggio:2015vxa,Gerchkovitz2017}
to test our predictions.  In the case of ${\cal N} = 4$ SYM with gauge group $SU(2)$ (or more properly
gauge algebra $\mathfrak{su}(2)$ in general), the exact result can be expressed in closed form,
and our asymptotic expansion for the logarithm of the two point function agrees precisely with the exact
result to the precision to which we have calculated, \it i.e. \rm up to and including the order $\log(\JJM)$ term
in ${\cal B}\ll n = \log (\ampname\ll n)$.  In the case of superconformal SQCD with $N\ll f = 2 N\ll c = 4$, we
compare our large-$\JJM$ expansion with the output of the recursion relations carried to $\JJM \simeq 60$,
with the $\JJM = 0$ expression approximated by the zero-instanton part of the $\Sfour$ partition function.  We find
precise numerical agreement for the two leading-order behaviors, and good agreement for the sub-subleading order
behavior, dictated by the $\a$-coefficient $\a = +{3\over 2}$, which predicts a value $-4$ for the 
LHS of the sum rule \rr{SQCDSumRuleFForDeltaEquals2Ord2} at large $n$.  Though it is not clear
we should expect the sum rule to approach $-4$ precisely for the zero-instanton approximation to the 
initial condition $Z\ll{\Sfour}$, the sum rule for $\t = i$ appears to asymptote to a value at most $\simeq -3.8$, to
within our numerical precision.  It would be desirable to have a robust theory of the error at large $\JJM$, given
an approximate initial condition for the recursion relation.\footnote{We thank Z. Komargodski for correspondence on this point.} It
may be a useful direction to study the recursion relations directly in a ${1\over \JJM}$ expansion, to understand to what
extent the large-$\JJM$ behavior is determined by initial condition $Z\ll{\Sfour}(\t,\tb)$ and to what extent it is guided by attractor phenomena
inherent to the recursion relations themselves.

In summary, we have shown that it is practical to use the large-$\JJM$ expansion as a bridge from the world of unbroken conformal symmetry, OPE data, and bootstraps, to the
world of the low-energy dynamics of the moduli space of vacua.

\section*{Acknowledgments}
The authors note helpful discussions
with Philip Argyres, Nozomu Kobayashi,  Markus Luty, Mauricio Romo, Vyacheslav Rychkov, Masataka Watanabe, and Alexander Zhiboedov.  
We thank Zohar Komargodski and Kyriakos Papadodimas
for reading the manuscript and for valuable comments.
We are particularly grateful to Daniel Jafferis
for early discussions on the relationship between
the large $R$-charge limit and the dynamics of
moduli space, as well as bringing refs. 
\cite{Baggio:2014ioa,Baggio:2014sna,Baggio:2015vxa, Gerchkovitz2017} to our attention as a possible check on the large-$\JJM$ expansion beyond the free-field approximation.  
The work of SH is supported by the World Premier International Research Center Initiative (WPI Initiative), MEXT, Japan; by the JSPS Program for Advancing Strategic International Networks to Accelerate the Circulation of Talented Researchers; and also supported in part by JSPS KAKENHI Grant Numbers JP22740153, JP26400242. SM   acknowledges the support by JSPS Research Fellowship for Young Scientists.    SH also thanks the Walter Burke Institute for Theoretical Physics at Caltech, the Stanford Institute for Theoretical Physics, and the Harvard Center for the Fundamental Laws of Nature, for hospitality while this work was in progress.

\appendix

\section{Normalizations and conventions}

\subsection{Massless scalar propagator}

Define the free massless complex scalar field $\phi\lrm{unit}$ 
to have kinetic term
\begin{align}
\begin{split}
{\cal L} = +1\cc
|\pp\phi\lrm{unit}|\sqd\ ,
\end{split}\label{UnitScalarKineticTermNormalizationDef}
\end{align}
in Euclidean signature, where $\pp\sqd \equiv \pp\ll\m\pp\ll\m$ and $|\pp\phi|\sqd
\equiv (\pp\ll\m \phi)(\pp\ll\m\phb)$.
The massless euclidean scalar propagator on $\IR\uu 4$
is defined as
\begin{align}
\begin{split}
\D\lrm{unit}(x,y) \equiv \left< \phi\lrm{unit}(x) \cc\cc
\phb\lrm{unit}(y) \right>
\ll{\IR\uu 4} \ .
\end{split}\label{MasslessScalarPropagatorUnitNormalization}
\end{align}
Given \rr{UnitScalarKineticTermNormalizationDef}, 
the Ward identity yields the equation of motion for the propagator
\begin{align}
\begin{split}
\pp\ll x\sqd \D\lrm {unit}(x,y) = \pp\ll y\sqd\D\lrm{unit}(x,y) = - \d\upp 4 (x - y)
\end{split}
\end{align}
so the propagator for the unit scalar has normalization
\begin{align}
\begin{split}
\D\lrm{unit}(x,y) = + (2\pi)\uu{-2} \cc |x - y|\uu{-2} 
\end{split}\label{PropagatorNormalizationForFieldWithUnitKineticTerm}
\end{align}
where we have used the identity
\begin{align}
\begin{split}
\partial^2_x \left|x - y\right|^{-2}= - (2\pi)^2 \delta^{(4)}(x-y)
\end{split}\label{ScalarPropagatorNormalization}
\end{align}
on $\IR\uu 4$.  More generally, for a massless
complex scalar field normalized as
\begin{align}
\begin{split}\label{ScalarKineticTermNormalizationDefGENERAL}
{\cal L}\ll{\bf M}\sqd = {\bf M}\sqd\cc
|\pp\ll \m \phi|\sqd\ ,
\end{split}
\end{align}
the scalar propagator is
\begin{align}
\begin{split}\label{MasslessScalarPropagatorGeneralNormalization}
\left<
 \phi(x) \cc
\phb(y) 
\right>
\ll{\IR\uu 4} = (2\pi)\uu{-2}\cc {\bf M}\uu{-2}\cc
|x - y|\uu{-2}
\end{split}
\end{align}
for any positive real ${\bf M}$.  In particular, for the $A$-field of the effective Abelian vector multiplet, whose kinetic term is
\rr{KineticTermForEffVMScalarA}, the two-point function is
\begin{align}
\begin{split}
\left< A(x) \cc\bar{A}(y) \right>\ll{\IR\uu 4} = {{g\sqd\lrm{eff}}\over{(2\pi)\sqd}} \cc 
|x - y|\uu{-2} = {1\over{\pi\cc{\rm Im}(\t)}}\cc |x - y|\uu{-2}.
\end{split}
\end{align}

\subsection{Geometry of the four-sphere}\label{SecFourSphere}

The four-sphere is a symmetric space, so its Riemann tensor satisfies
\begin{align}
\begin{split}
R_{abcd} = {1\over{r\sqd}} (g_{ac} g_{bd} - g_{ad} g_{bc}) \ .
\end{split}
\end{align}
So for a general $D$-dimensional sphere we have
\begin{align}
\begin{split}
R_{ac} = g^{bd} R_{abcd} = {1\over{r\sqd}}\cc (D-1) g_{ac}
\ , \qquad
R = g^{ac} R_{ac} =  {{D(D-1)}\over{r\sqd}}\ .
\end{split}
\end{align}
Now let us calculate the Euler density, according to Komargodski-Schwimmer's normalization convention
\rr{KomargodskiSchwimmerEulerDensityNormalization}.  The square of the Riemann tensor is
\begin{align}
\begin{split}
R_{abcd}^2\equiv g\uu{aa\pr}g\uu{bb\pr}g\uu{cc\pr}g\uu{dd\pr}\cc R\ll{abcd} R\ll{a\pr b\pr c\pr d\pr}  = 2\cc {{D(D-1)}\over{ r^4}}
\label{RiemannSquaredSphere}
\end{split}
\end{align}
and the squares of the Ricci tensor and Ricci scalar are
\begin{align}\label{RicciTensorSquaredSphere}
\begin{split}
R_{ab} =  {{D-1}\over{r\sqd}} \cc g_{ab}
\ ,  \qquad
R_{ab}^2 \equiv g^{ac} g^{bd} R_{ab} R_{cd} = {{D(D-1)\sqd}\over{r\uu 4}}\ .
\end{split}
\end{align}
The Ricci scalar and its square are
\begin{align}
\begin{split}
R = {{D(D-1)}\over{r\sqd}}\ , \qquad
R\sqd = {{D\sqd(D-1)\sqd}\over{r\uu 4}}\ .
\end{split}\label{RicciScalarSphere}
\end{align}
The case of interest to us is $D=4$, in which
\begin{align}
\begin{split}
(R_{abcd}^2)\ll{D = 4} = {{24}\over{r\uu 4}}\ ,
\qquad
(R\ll{ab}\sqd)\ll{D = 4} = {{36}\over{r\uu 4}}\ ,
\qquad
(R^2)\ll{D = 4} = {{144}\over{r\uu 4}}\ .
\end{split}
\end{align}
Komargodski-Schwimmer's normalization of the Euler density, in their equation (A.4), is
\begin{align}
\begin{split}
E_4\uu{\rm [KS]}\equiv R_{abcd}^2 - 4 R_{ab}^2 + R^2\ ,
\end{split}
\end{align}
which for the four-sphere of radius $r$, is given by
\begin{align}
\begin{split}
E_4 \uu{\rm [KS]} = {{24}\over{r^4}}\ .
\end{split}
\end{align}

\subsection{Conventions and values for the $a$-anomaly coefficient }\label{AnomalyConventionsAndValues}

In this part of the Appendix, we compare two conventions for the normalization of the $a$-anomaly coefficient (also the
$c$-anomaly coefficient), and we give values for the anomaly in various ${\cal N} = 2$ SCFT of interest.  We also
give a definition of the $\a$-coefficient that is independent of the normalization of the $a$-anomaly.

\heading{Translation between Weyl-anomaly normalization conventions in \cite{Komargodski:2011vj}  vs. \cite{Anselmi:1997ys}}

The $a$- and $c$-anomalies are normalized differently in different parts of the literature.  We can match by
comparing anomalies for a given physical system across conventions.  The simplest case is a scalar field.

In \cite{Komargodski:2011vj} the anomalies are normalized so that the contributions of a single real massless scalar field, are
\begin{align}
\begin{split}
a\uu{\rm [KS]}_{\text{real massless scalar}} =\frac{1}{90 (8\pi)^2}\ , \qquad
c\uu{\rm [KS]}_{\text{real massless scalar}}  =\frac{1}{30 (8\pi)^2}\ .
\end{split}\label{AnomalyOfAScalarKS}
\end{align}
This normalization is given below equation (A.6) of  \cite{Komargodski:2011vj}.
In \cite{Anselmi:1997ys}, the authors give the anomalies of a single real massless scalar field, as 
\begin{align}
\begin{split}
\label{AnomalyOfAScalarAEFJ}
a\uu{\rm [AEFJ]}_{\text{real massless scalar}} =\frac{1}{360}\ , \qquad
c\uu{\rm [AEFJ]}_{\text{real massless scalar}}  =\frac{1}{120}\ .
\end{split}
\end{align}
The relation between the two normalizations is therefore
\begin{align}
\begin{split}
a\uu{\rm [KS]} = {1\over{16\pi\sqd}} \cc a\uu{\rm [AEFJ]}\ , 
\qquad
c\uu{\rm [KS]} = {1\over{16\pi\sqd}} \cc c\uu{\rm [AEFJ]}.
\end{split}
\end{align}
In the body of the paper we indicate our conventions to avoid ambiguity, but we shall use the convention
of \cite{Anselmi:1997ys}, since it is normalized such that the anomalies of free fields, and of all ${\cal N} = 2$ SCFT,
are rational numbers.

\heading{Values of the anomaly coefficient in various ${\cal N} = 2$ SCFT in $D=4$}

We have defined the
exponent $\a$, which appears in the factor $\JJM\uu \a$
in the asymptotic formula for the two-point function, in terms of the $a$ coefficient in the Weyl anomaly.
The Weyl anomaly does not have a universally used normalization in the literature.  So
in order to find actual values 
for $\a$, we need to use some particular conventions.

The $a$-coefficients for many Lagrangian and non-Lagrangian theories, have been given in \it e.g. \rm 
\cite{Anselmi:1997ys, Shapere:2008zf}, and we collect the relevant results here.
Those authors normalize the $a$-coefficient according to the widely-used
convention in \cite{Anselmi:1997ys}, in which we have
\begin{align}
\begin{split}
a\ll{\rm favm}\uu{\rm [AEFJ]} = {5\over{24}},
\quad
a\ll{\rm fhm}\uu{\rm [AEFJ]} = {1\over{24}}\ ,
\quad
a\ll{{\rm vm}\ll G}\uu{\rm [AEFJ]} = {5\over{24}} \cc
\dim(G)\ ,
\quad
a\ll{{\rm hm}\ll R}\uu{\rm [AEFJ]} = {1\over{24}} \cc
\dim_{\mathbb C}(R)\ .
\end{split}
\end{align}

\heading{Value of the $a$-coefficient for ${\cal N} = 4$ SYM}
Organizing into ${\cal N} = 4$ vectormultiplets, we have
\begin{align}
\begin{split}
a\ll{{\cal N}=4\rm~favm}\uu{\rm [AEFJ]} = 
a\ll{\rm favm}\uu{\rm [AEFJ]} + a\ll{\rm fhm}\uu{\rm [AEFJ]} = \frac14\ ,
\qquad
a\uu{\rm [AEFJ]}\ll{{\cal N}=4\rm ~vm\ll G} = {1\over 4}\cc\dim(G)\ .
\end{split}
\end{align}
The ${\cal N} = 4$ theory with gauge group $G$ has
microscopic $a$-coefficient
\begin{align}
\begin{split}
a\uu{\rm [AEFJ]}\lrm{CFT,~{\cal N} =4} = {1\over 4}\cc \dim(G)
\end{split}
\end{align}
and its moduli space effective theory has 
\begin{align}
\begin{split}
a\uu{\rm [AEFJ]}\lrm{EFT,~{\cal N} =4} = {1\over 4}\cc \operatorname{\mathop{rank}}(G)
\end{split}
\end{align}
so
\begin{align}
\begin{split}
\Delta a \uu{\rm [AEFJ]}\lrm{{\cal N} =4} = 
{1\over 4}\cc\left ( \cc\dim(G) -  \operatorname{\mathop{rank}}(G) \cc \right )\ .
\end{split}
\end{align}
For $SU(N\ll c)$ gauge group, we have
\begin{align}
\begin{split}
\Delta a \uu{\rm [AEFJ]}\ll{{\cal N} =4,~G = SU(N\ll c)} = 
 {1\over 4}(N\ll c\sqd - N\ll c)\ .
\end{split}
\end{align}
In particular, for $N\ll c = 2$ we have
\begin{align}
\begin{split}
\Delta a \uu{\rm [AEFJ]}\ll{{\cal N} =4,~G = SU(2)} = 
\frac12.
\end{split}
\end{align}

\heading{Value of the $a$-coefficient for superconformal ${\cal N} = 2$ SQCD}
For ${\cal N} = 2$ SQCD with gauge group $SU(N\ll c)$ and $N\ll f$ fundamental
flavors at weak coupling, we have
\begin{align}
\begin{split}
a\uu{\rm [AEFJ]}\lrm{UV,~SQCD} = {5\over {24}}\cc \dim(G) + {1\over{24}}\cc \dim\ll{\mathbb C}(R)
= {5\over{24}}(N\ll c\sqd - 1) + {1\over{24}}\cc N\ll f N\ll c\ .
\end{split}
\end{align}
In the superconformal case, $N\ll f = 2 N\ll c$ we have
\begin{align}
\begin{split}
a\uu{\rm [AEFJ]}\lrm{CFT,~SCQCD} = {5\over {24}}\cc \dim(G) + {1\over{24}}\cc\dim\ll{\mathbb C}(R)
= {5\over{24}}(N\ll c\sqd - 1) + {1\over{12}}\cc  N\ll c\sqd
 = {7\over{24}}\cc N\ll c\sqd - {5\over{24}}\ .
\end{split}
\end{align}
The moduli space effective theory consists of $\operatorname{\mathop{rank}}(G)$
free abelian vector multiplets and no hypers, so we have
\begin{align}
\begin{split}
a\uu{\rm [AEFJ]}\lrm{EFT,~SCQCD} = {5\over {24}}\cc \operatorname{\mathop{rank}}(G) \ ,
\end{split}
\end{align}
which for $G = SU(N)$, is
\begin{align}
\begin{split}
a\uu{\rm [AEFJ]}\lrm{EFT,~SCQCD} = {5\over {24}}\cc (N\ll c - 1)\ .
\end{split}
\end{align}
So the difference in central charge is
\begin{align}
\begin{split}
\Delta a\uu{\rm [AEFJ]}\lrm{SCQCD} =
a\uu{\rm [AEFJ]}\lrm{CFT,~SCQCD}  -
a\uu{\rm [AEFJ]}\lrm{EFT,~SCQCD} 
= {1\over{24}} \cc \left ( \cc 7 N\ll c\sqd - 5 N\ll c  \cc\right )\ .
\end{split}
\end{align}

\heading{Value of the $\a$-coefficient for ${\cal N} = 4$ SYM}
So for ${\cal N} = 4$ we have
\begin{align}
\begin{split}
\a\ll{{\cal N} = 4} = 
\hh\cc (N\ll c\sqd - N\ll c)\ .
\end{split}
\end{align}
In particular, for $G = SU(2)$ we have
\begin{align}
\begin{split}\label{NEqualsFourAlpheCoefficient}
\a\ll{{\cal N} = 4,~G = SU(2)} = +1\ . 
\end{split}
\end{align}

\heading{Value of the $\a$-coefficient for ${\cal N} = 2$ superconformal SQCD}

For superconformal QCD (that is, $N\ll f = 2 N\ll c$), we 
have
\begin{align}
\begin{split}
\a\ll{{\rm SCQCD}}  
 = {1\over{12}}\cc 
\left ( \cc 7 N\ll c\sqd - 5 N\ll c  \cc\right )\ ,
\end{split}
\end{align}
and in particular for $G = SU(2)$ with $N\ll f = 2 N\ll c = 4$,
we have
\begin{align}
\begin{split}
\a\ll{{\rm SCQCD},~G=SU(2)} =\frac32
\label{SQCDAlphaCoefficientValue}
\end{split}
\end{align}

\heading{Convention-independent formula for the $\a$-coefficient}

We would like to define the $\a$-coefficient in a convention-independent way, as a ratio of
$a$-anomalies.  Our convention-independent formula is:
\begin{align}
\begin{split}\label{ConventionIndependentFormulaForAlphaRECAP}
\a = {5\over {12}}\cc
{{a\lrm{CFT} - a\lrm{EFT}}\over{a\lrm{favm}}}\ ,
\end{split}
\end{align}
where $a\lrm{favm}$ is the unit of $a$-anomaly contribution
carried by a free ${\cal N} = 2$ vector multiplet for
a $U(1)$ gauge group.
In order to actually compute the value of $\a$ for some
theories of interest, we must pick an actual
normalization convention.  The value of $\a$ in the \cite{Anselmi:1997ys} convention is
\begin{align}
\begin{split}
\a = 2 \cc \left ( \cc a\uu{\rm [AEFJ]}\lrm{CFT} - a\uu{\rm [AEFJ]}\lrm{EFT} \cc \right )\ ,
\end{split}
\end{align}
and in the \cite{Komargodski:2011vj} convention it is
\begin{align}
\begin{split}
\a = {1\over{8\pi\sqd}}\cc \left ( \cc a\uu{\rm [KS]}\lrm{CFT} - a\uu{\rm [KS]}\lrm{EFT} \cc \right )\ .
\end{split}
\end{align}


\bibliographystyle{utphys-edited} 
\bibliography{references-susy-two-point}

\providecommand{\href}[2]{#2}\begingroup\raggedright\begin{thebibliography}{10}

\bibitem{Hellerman:2017veg}
S.~Hellerman, S.~Maeda, and M.~Watanabe, {\textquotedbl}{Operator Dimensions
  from Moduli},{\textquotedbl}
  \href{http://dx.doi.org/10.1007/JHEP10(2017)089}{{\em JHEP} {\bfseries 10}
  (2017) 89},
\href{http://arxiv.org/abs/1706.05743}{{\ttfamily arXiv:1706.05743 [hep-th]}}.

\bibitem{Baggio:2014ioa}
M.~Baggio, V.~Niarchos, and K.~Papadodimas, {\textquotedbl}{$tt^{*}$ equations,
  localization and exact chiral rings in 4d $ \mathcal{N} $ =2
  SCFTs},{\textquotedbl} \href{http://dx.doi.org/10.1007/JHEP02(2015)122}{{\em
  JHEP} {\bfseries 02} (2015) 122},
\href{http://arxiv.org/abs/1409.4212}{{\ttfamily arXiv:1409.4212 [hep-th]}}.

\bibitem{Baggio:2014sna}
M.~Baggio, V.~Niarchos, and K.~Papadodimas, {\textquotedbl}{Exact correlation
  functions in $SU(2)$ $\mathcal N=2$ superconformal QCD},{\textquotedbl}
  \href{http://dx.doi.org/10.1103/PhysRevLett.113.251601}{{\em Phys. Rev.
  Lett.} {\bfseries 113} no.~25, (2014) 251601},
\href{http://arxiv.org/abs/1409.4217}{{\ttfamily arXiv:1409.4217 [hep-th]}}.

\bibitem{Baggio:2015vxa}
M.~Baggio, V.~Niarchos, and K.~Papadodimas, {\textquotedbl}{On exact
  correlation functions in $SU(N)$ $ \mathcal{N}=2 $ superconformal
  QCD},{\textquotedbl} \href{http://dx.doi.org/10.1007/JHEP11(2015)198}{{\em
  JHEP} {\bfseries 11} (2015) 198},
\href{http://arxiv.org/abs/1508.03077}{{\ttfamily arXiv:1508.03077 [hep-th]}}.

\bibitem{Gerchkovitz2017}
E.~Gerchkovitz, J.~Gomis, N.~Ishtiaque, A.~Karasik, Z.~Komargodski, and S.~S.
  Pufu, {\textquotedbl}{Correlation Functions of Coulomb Branch
  Operators},{\textquotedbl}
  \href{http://dx.doi.org/10.1007/JHEP01(2017)103}{{\em JHEP} {\bfseries 01}
  (2017) 103},
\href{http://arxiv.org/abs/1602.05971}{{\ttfamily arXiv:1602.05971 [hep-th]}}.

\bibitem{Argyres:2015ffa}
P.~Argyres, M.~Lotito, Y.~L{\"u}, and M.~Martone, {\textquotedbl}{Geometric
  constraints on the space of N=2 SCFTs I: physical constraints on relevant
  deformations},{\textquotedbl}
\href{http://arxiv.org/abs/1505.04814}{{\ttfamily arXiv:1505.04814 [hep-th]}}.

\bibitem{Argyres:2015gha}
P.~C. Argyres, M.~Lotito, Y.~L{\"u}, and M.~Martone, {\textquotedbl}{Geometric
  constraints on the space of N=2 SCFTs II: Construction of special K\"ahler
  geometries and RG flows},{\textquotedbl}
\href{http://arxiv.org/abs/1601.00011}{{\ttfamily arXiv:1601.00011 [hep-th]}}.

\bibitem{Argyres:2016xmc}
P.~Argyres, M.~Lotito, Y.~L{\"u}, and M.~Martone, {\textquotedbl}{Geometric
  constraints on the space of N=2 SCFTs III: enhanced Coulomb branches and
  central charges},{\textquotedbl}
\href{http://arxiv.org/abs/1609.04404}{{\ttfamily arXiv:1609.04404 [hep-th]}}.

\bibitem{Argyres:2016xua}
P.~C. Argyres, M.~Lotito, Y.~L{\"u}, and M.~Martone, {\textquotedbl}{Expanding
  the landscape of $ \mathcal{N} $ = 2 rank 1 SCFTs},{\textquotedbl}
  \href{http://dx.doi.org/10.1007/JHEP05(2016)088}{{\em JHEP} {\bfseries 05}
  (2016) 088},
\href{http://arxiv.org/abs/1602.02764}{{\ttfamily arXiv:1602.02764 [hep-th]}}.

\bibitem{Hellerman:2015nra}
S.~Hellerman, D.~Orlando, S.~Reffert, and M.~Watanabe, {\textquotedbl}{On the
  CFT Operator Spectrum at Large Global Charge},{\textquotedbl}
  \href{http://dx.doi.org/10.1007/JHEP12(2015)071}{{\em JHEP} {\bfseries 12}
  (2015) 071},
\href{http://arxiv.org/abs/1505.01537}{{\ttfamily arXiv:1505.01537 [hep-th]}}.

\bibitem{Alvarez-Gaume:2016vff}
L.~Alvarez-Gaume, O.~Loukas, D.~Orlando, and S.~Reffert,
  {\textquotedbl}{Compensating strong coupling with large
  charge},{\textquotedbl} \href{http://dx.doi.org/10.1007/JHEP04(2017)059}{{\em
  JHEP} {\bfseries 04} (2017) 059},
\href{http://arxiv.org/abs/1610.04495}{{\ttfamily arXiv:1610.04495 [hep-th]}}.

\bibitem{Monin:2016jmo}
A.~Monin, D.~Pirtskhalava, R.~Rattazzi, and F.~K. Seibold,
  {\textquotedbl}{Semiclassics, Goldstone Bosons and CFT data},{\textquotedbl}
  \href{http://dx.doi.org/10.1007/JHEP06(2017)011}{{\em JHEP} {\bfseries 06}
  (2017) 011},
\href{http://arxiv.org/abs/1611.02912}{{\ttfamily arXiv:1611.02912 [hep-th]}}.

\bibitem{Loukas:2016ckj}
O.~Loukas, {\textquotedbl}{Abelian scalar theory at large global
  charge},{\textquotedbl} \href{http://dx.doi.org/10.1002/prop.201700028}{{\em
  Fortsch. Phys.} {\bfseries 65} no.~9, (2017) 1700028},
\href{http://arxiv.org/abs/1612.08985}{{\ttfamily arXiv:1612.08985 [hep-th]}}.

\bibitem{Hellerman:2017efx}
S.~Hellerman, N.~Kobayashi, S.~Maeda, and M.~Watanabe, {\textquotedbl}{A Note
  on Inhomogeneous Ground States at Large Global Charge},{\textquotedbl}
\href{http://arxiv.org/abs/1705.05825}{{\ttfamily arXiv:1705.05825 [hep-th]}}.

\bibitem{Loukas:2017lof}
O.~Loukas, D.~Orlando, and S.~Reffert, {\textquotedbl}{Matrix models at large
  charge},{\textquotedbl} \href{http://dx.doi.org/10.1007/JHEP10(2017)085}{{\em
  JHEP} {\bfseries 10} (2017) 085},
\href{http://arxiv.org/abs/1707.00710}{{\ttfamily arXiv:1707.00710 [hep-th]}}.

\bibitem{Banerjee:2017fcx}
D.~Banerjee, S.~Chandrasekharan, and D.~Orlando, {\textquotedbl}{Conformal
  dimensions via large charge expansion},{\textquotedbl}
\href{http://arxiv.org/abs/1707.00711}{{\ttfamily arXiv:1707.00711 [hep-lat]}}.

\bibitem{Fitzpatrick:2012yx}
A.~L. Fitzpatrick, J.~Kaplan, D.~Poland, and D.~Simmons-Duffin,
  {\textquotedbl}{The Analytic Bootstrap and AdS Superhorizon
  Locality},{\textquotedbl}
  \href{http://dx.doi.org/10.1007/JHEP12(2013)004}{{\em JHEP} {\bfseries 12}
  (2013) 004},
\href{http://arxiv.org/abs/1212.3616}{{\ttfamily arXiv:1212.3616 [hep-th]}}.

\bibitem{Komargodski:2012ek}
Z.~Komargodski and A.~Zhiboedov, {\textquotedbl}{Convexity and Liberation at
  Large Spin},{\textquotedbl}
  \href{http://dx.doi.org/10.1007/JHEP11(2013)140}{{\em JHEP} {\bfseries 11}
  (2013) 140},
\href{http://arxiv.org/abs/1212.4103}{{\ttfamily arXiv:1212.4103 [hep-th]}}.

\bibitem{Li:2015itl}
D.~Li, D.~Meltzer, and D.~Poland, {\textquotedbl}{Conformal Collider Physics
  from the Lightcone Bootstrap},{\textquotedbl}
  \href{http://dx.doi.org/10.1007/JHEP02(2016)143}{{\em JHEP} {\bfseries 02}
  (2016) 143},
\href{http://arxiv.org/abs/1511.08025}{{\ttfamily arXiv:1511.08025 [hep-th]}}.

\bibitem{Li:2015rfa}
D.~Li, D.~Meltzer, and D.~Poland, {\textquotedbl}{Non-Abelian Binding Energies
  from the Lightcone Bootstrap},{\textquotedbl}
  \href{http://dx.doi.org/10.1007/JHEP02(2016)149}{{\em JHEP} {\bfseries 02}
  (2016) 149},
\href{http://arxiv.org/abs/1510.07044}{{\ttfamily arXiv:1510.07044 [hep-th]}}.

\bibitem{Kaviraj:2015cxa}
A.~Kaviraj, K.~Sen, and A.~Sinha, {\textquotedbl}{Analytic bootstrap at large
  spin},{\textquotedbl} \href{http://dx.doi.org/10.1007/JHEP11(2015)083}{{\em
  JHEP} {\bfseries 11} (2015) 083},
\href{http://arxiv.org/abs/1502.01437}{{\ttfamily arXiv:1502.01437 [hep-th]}}.

\bibitem{Kaviraj:2015xsa}
A.~Kaviraj, K.~Sen, and A.~Sinha, {\textquotedbl}{Universal anomalous
  dimensions at large spin and large twist},{\textquotedbl}
  \href{http://dx.doi.org/10.1007/JHEP07(2015)026}{{\em JHEP} {\bfseries 07}
  (2015) 026},
\href{http://arxiv.org/abs/1504.00772}{{\ttfamily arXiv:1504.00772 [hep-th]}}.

\bibitem{Dey:2017fab}
P.~Dey, K.~Ghosh, and A.~Sinha, {\textquotedbl}{Simplifying large spin
  bootstrap in Mellin space},{\textquotedbl}
\href{http://arxiv.org/abs/1709.06110}{{\ttfamily arXiv:1709.06110 [hep-th]}}.

\bibitem{Alday:2015ewa}
L.~F. Alday and A.~Zhiboedov, {\textquotedbl}{An Algebraic Approach to the
  Analytic Bootstrap},{\textquotedbl}
  \href{http://dx.doi.org/10.1007/JHEP04(2017)157}{{\em JHEP} {\bfseries 04}
  (2017) 157},
\href{http://arxiv.org/abs/1510.08091}{{\ttfamily arXiv:1510.08091 [hep-th]}}.

\bibitem{Alday:2016njk}
L.~F. Alday, {\textquotedbl}{Large Spin Perturbation Theory},{\textquotedbl}
  \href{http://dx.doi.org/10.1103/PhysRevLett.119.111601}{{\em Phys. Rev.
  Lett.} {\bfseries 119} no.~11, (2017) 111601},
\href{http://arxiv.org/abs/1611.01500}{{\ttfamily arXiv:1611.01500 [hep-th]}}.

\bibitem{Alday:2016jfr}
L.~F. Alday, {\textquotedbl}{Solving CFTs with Weakly Broken Higher Spin
  Symmetry},{\textquotedbl}
\href{http://arxiv.org/abs/1612.00696}{{\ttfamily arXiv:1612.00696 [hep-th]}}.

\bibitem{Hellerman:2016hnf}
S.~Hellerman and I.~Swanson, {\textquotedbl}{Boundary Operators in Effective
  String Theory},{\textquotedbl}
  \href{http://dx.doi.org/10.1007/JHEP04(2017)085}{{\em JHEP} {\bfseries 04}
  (2017) 085},
\href{http://arxiv.org/abs/1609.01736}{{\ttfamily arXiv:1609.01736 [hep-th]}}.

\bibitem{Hellerman:2014cba}
S.~Hellerman, S.~Maeda, J.~Maltz, and I.~Swanson, {\textquotedbl}{Effective
  String Theory Simplified},{\textquotedbl}
  \href{http://dx.doi.org/10.1007/JHEP09(2014)183}{{\em JHEP} {\bfseries 09}
  (2014) 183},
\href{http://arxiv.org/abs/1405.6197}{{\ttfamily arXiv:1405.6197 [hep-th]}}.

\bibitem{Hellerman:2013kba}
S.~Hellerman and I.~Swanson, {\textquotedbl}{String Theory of the Regge
  Intercept},{\textquotedbl}
  \href{http://dx.doi.org/10.1103/PhysRevLett.114.111601}{{\em Phys. Rev.
  Lett.} {\bfseries 114} no.~11, (2015) 111601},
\href{http://arxiv.org/abs/1312.0999}{{\ttfamily arXiv:1312.0999 [hep-th]}}.

\bibitem{Costa:2012cb}
M.~S. Costa, V.~Goncalves, and J.~Penedones, {\textquotedbl}{Conformal Regge
  theory},{\textquotedbl} \href{http://dx.doi.org/10.1007/JHEP12(2012)091}{{\em
  JHEP} {\bfseries 12} (2012) 091},
\href{http://arxiv.org/abs/1209.4355}{{\ttfamily arXiv:1209.4355 [hep-th]}}.

\bibitem{Caron-Huot:2016icg}
S.~Caron-Huot, Z.~Komargodski, A.~Sever, and A.~Zhiboedov,
  {\textquotedbl}{Strings from Massive Higher Spins: The Asymptotic Uniqueness
  of the Veneziano Amplitude},{\textquotedbl}
  \href{http://dx.doi.org/10.1007/JHEP10(2017)026}{{\em JHEP} {\bfseries 10}
  (2017) 026},
\href{http://arxiv.org/abs/1607.04253}{{\ttfamily arXiv:1607.04253 [hep-th]}}.

\bibitem{Sever:2017ylk}
A.~Sever and A.~Zhiboedov, {\textquotedbl}{On Fine Structure of Strings: The
  Universal Correction to the Veneziano Amplitude},{\textquotedbl}
\href{http://arxiv.org/abs/1707.05270}{{\ttfamily arXiv:1707.05270 [hep-th]}}.

\bibitem{Son:2005rv}
D.~T. Son and M.~Wingate, {\textquotedbl}{General coordinate invariance and
  conformal invariance in nonrelativistic physics: Unitary Fermi
  gas},{\textquotedbl} \href{http://dx.doi.org/10.1016/j.aop.2005.11.001}{{\em
  Annals Phys.} {\bfseries 321} (2006) 197--224},
\href{http://arxiv.org/abs/cond-mat/0509786}{{\ttfamily arXiv:cond-mat/0509786
  [cond-mat]}}.

\bibitem{Rattazzi:2008pe}
R.~Rattazzi, V.~S. Rychkov, E.~Tonni, and A.~Vichi, {\textquotedbl}{Bounding
  scalar operator dimensions in 4D CFT},{\textquotedbl}
  \href{http://dx.doi.org/10.1088/1126-6708/2008/12/031}{{\em JHEP} {\bfseries
  12} (2008) 031},
\href{http://arxiv.org/abs/0807.0004}{{\ttfamily arXiv:0807.0004 [hep-th]}}.

\bibitem{El-Showk:2014dwa}
S.~El-Showk, M.~F. Paulos, D.~Poland, S.~Rychkov, D.~Simmons-Duffin, and
  A.~Vichi, {\textquotedbl}{Solving the 3d Ising Model with the Conformal
  Bootstrap II. $c$-Minimization and Precise Critical
  Exponents},{\textquotedbl}
  \href{http://dx.doi.org/10.1007/s10955-014-1042-7}{{\em J. Stat. Phys.}
  {\bfseries 157} (2014) 869},
\href{http://arxiv.org/abs/1403.4545}{{\ttfamily arXiv:1403.4545 [hep-th]}}.

\bibitem{ElShowk:2012ht}
S.~El-Showk, M.~F. Paulos, D.~Poland, S.~Rychkov, D.~Simmons-Duffin, and
  A.~Vichi, {\textquotedbl}{Solving the 3D Ising Model with the Conformal
  Bootstrap},{\textquotedbl}
  \href{http://dx.doi.org/10.1103/PhysRevD.86.025022}{{\em Phys. Rev.}
  {\bfseries D86} (2012) 025022},
\href{http://arxiv.org/abs/1203.6064}{{\ttfamily arXiv:1203.6064 [hep-th]}}.

\bibitem{Rychkov:2016iqz}
S.~Rychkov, {\textquotedbl}{EPFL Lectures on Conformal Field Theory in $D\geq
  3$ Dimensions},{\textquotedbl}
  \href{http://dx.doi.org/10.1007/978-3-319-43626-5}{{\em SpringerBriefs in
  Physics} (2016) },
\href{http://arxiv.org/abs/1601.05000}{{\ttfamily arXiv:1601.05000 [hep-th]}}.

\bibitem{Simmons-Duffin:2016gjk}
D.~Simmons-Duffin, {\textquotedbl}{The Conformal Bootstrap},{\textquotedbl}
  \href{http://dx.doi.org/10.1142/9789813149441_0001}{{\em {Proceedings,
  Theoretical Advanced Study Institute in Elementary Particle Physics: New
  Frontiers in Fields and Strings (TASI 2015): Boulder, CO, USA, June 1-26,
  2015}} (2017) 1--74},
\href{http://arxiv.org/abs/1602.07982}{{\ttfamily arXiv:1602.07982 [hep-th]}}.

\bibitem{PhysRevE.50.888}
M.~Srednicki, {\textquotedbl}Chaos and quantum thermalization,{\textquotedbl}
  \href{http://dx.doi.org/10.1103/PhysRevE.50.888}{{\em Phys. Rev.} {\bfseries
  E50} (1994) 888--901},
  \href{http://arxiv.org/abs/cond-mat/9403051}{{\ttfamily
  arXiv:cond-mat/9403051 [cond-mat]}}.

\bibitem{Jafferis:2017zna}
D.~Jafferis, B.~Mukhametzhanov, and A.~Zhiboedov, {\textquotedbl}{Conformal
  Bootstrap At Large Charge},{\textquotedbl}
\href{http://arxiv.org/abs/1710.11161}{{\ttfamily arXiv:1710.11161 [hep-th]}}.

\bibitem{sashatalk}
A.~Zhiboedov, {\textquotedbl}{Conformal Bootstrap At Large
  Charge},{\textquotedbl} {MS seminar at Kavli Institute for the Physics and
  Mathematics of the Universe}.
\newblock Oct., 2017.
\newblock \url{http://research.ipmu.jp/seminar/?seminar_id=1939}.

\bibitem{Rodriguez-Gomez:2016ijh}
D.~Rodriguez-Gomez and J.~G. Russo, {\textquotedbl}{Large $N$ Correlation
  Functions in Superconformal Field Theories},{\textquotedbl}
  \href{http://dx.doi.org/10.1007/JHEP06(2016)109}{{\em JHEP} {\bfseries 06}
  (2016) 109},
\href{http://arxiv.org/abs/1604.07416}{{\ttfamily arXiv:1604.07416 [hep-th]}}.

\bibitem{Dedushenko:2016jxl}
M.~Dedushenko, S.~S. Pufu, and R.~Yacoby, {\textquotedbl}{A one-dimensional
  theory for Higgs branch operators},{\textquotedbl}
\href{http://arxiv.org/abs/1610.00740}{{\ttfamily arXiv:1610.00740 [hep-th]}}.

\bibitem{Baggio:2016skg}
M.~Baggio, V.~Niarchos, K.~Papadodimas, and G.~Vos, {\textquotedbl}{Large-N
  correlation functions in $ \mathcal{N} $ = 2 superconformal
  QCD},{\textquotedbl} \href{http://dx.doi.org/10.1007/JHEP01(2017)101}{{\em
  JHEP} {\bfseries 01} (2017) 101},
\href{http://arxiv.org/abs/1610.07612}{{\ttfamily arXiv:1610.07612 [hep-th]}}.

\bibitem{Pini:2017ouj}
A.~Pini, D.~Rodriguez-Gomez, and J.~G. Russo, {\textquotedbl}{Large $N$
  correlation functions in $ \mathcal{N}=$ 2 superconformal
  quivers},{\textquotedbl}
  \href{http://dx.doi.org/10.1007/JHEP08(2017)066}{{\em JHEP} {\bfseries 08}
  (2017) 066},
\href{http://arxiv.org/abs/1701.02315}{{\ttfamily arXiv:1701.02315 [hep-th]}}.

\bibitem{Baggio:2017aww}
M.~Baggio, V.~Niarchos, and K.~Papadodimas, {\textquotedbl}{Aspects of Berry
  phase in QFT},{\textquotedbl}
  \href{http://dx.doi.org/10.1007/JHEP04(2017)062}{{\em JHEP} {\bfseries 04}
  (2017) 062},
\href{http://arxiv.org/abs/1701.05587}{{\ttfamily arXiv:1701.05587 [hep-th]}}.

\bibitem{Billo:2017glv}
M.~Billo, F.~Fucito, A.~Lerda, J.~F. Morales, {\relax Ya}.~S. Stanev, and
  C.~Wen, {\textquotedbl}{Two-point Correlators in ${\mathcal{N}=2}$ Gauge
  Theories},{\textquotedbl}
\href{http://arxiv.org/abs/1705.02909}{{\ttfamily arXiv:1705.02909 [hep-th]}}.

\bibitem{Agmon:2017lga}
N.~B. Agmon, S.~M. Chester, and S.~S. Pufu, {\textquotedbl}{A New Duality
  Between $\mathcal{N}=8$ Superconformal Field Theories in Three
  Dimensions},{\textquotedbl}
\href{http://arxiv.org/abs/1708.07861}{{\ttfamily arXiv:1708.07861 [hep-th]}}.

\bibitem{Papadodimas:2009eu}
K.~Papadodimas, {\textquotedbl}{Topological Anti-Topological Fusion in
  Four-Dimensional Superconformal Field Theories},{\textquotedbl}
  \href{http://dx.doi.org/10.1007/JHEP08(2010)118}{{\em JHEP} {\bfseries 08}
  (2010) 118},
\href{http://arxiv.org/abs/0910.4963}{{\ttfamily arXiv:0910.4963 [hep-th]}}.

\bibitem{Anselmi:1997ys}
D.~Anselmi, J.~Erlich, D.~Z. Freedman, and A.~A. Johansen,
  {\textquotedbl}{Positivity Constraints on Anomalies in Supersymmetric Gauge
  Theories},{\textquotedbl}
  \href{http://dx.doi.org/10.1103/PhysRevD.57.7570}{{\em Phys. Rev.} {\bfseries
  D57} (1998) 7570--7588},
\href{http://arxiv.org/abs/hep-th/9711035}{{\ttfamily arXiv:hep-th/9711035
  [hep-th]}}.

\bibitem{Argyres:2003tg}
P.~C. Argyres, A.~M. Awad, G.~A. Braun, and F.~P. Esposito,
  {\textquotedbl}{Higher derivative terms in $N=2$ supersymmetric effective
  actions},{\textquotedbl}
  \href{http://dx.doi.org/10.1088/1126-6708/2003/07/060}{{\em JHEP} {\bfseries
  07} (2003) 060},
\href{http://arxiv.org/abs/hep-th/0306118}{{\ttfamily arXiv:hep-th/0306118
  [hep-th]}}.

\bibitem{Argyres:2004kg}
P.~C. Argyres, A.~M. Awad, G.~A. Braun, and F.~P. Esposito,
  {\textquotedbl}{Higher derivative terms in $N=2$ SUSY effective
  actions},{\textquotedbl}
  \href{http://dx.doi.org/10.1142/9789812702340_0034}{{\em {Proceedings, 3rd
  International Symposium on Quantum theory and symmetries (QTS3): Cincinnati,
  USA, September 10-14, 2003}} (2004) 287--293},
\href{http://arxiv.org/abs/hep-th/0402203}{{\ttfamily arXiv:hep-th/0402203
  [hep-th]}}.

\bibitem{Argyres:2004yp}
P.~C. Argyres, A.~M. Awad, G.~A. Braun, and F.~P. Esposito, {\textquotedbl}{On
  superspace Chern-Simons-like terms},{\textquotedbl}
  \href{http://dx.doi.org/10.1088/1126-6708/2005/02/006}{{\em JHEP} {\bfseries
  02} (2005) 006},
\href{http://arxiv.org/abs/hep-th/0411081}{{\ttfamily arXiv:hep-th/0411081
  [hep-th]}}.

\bibitem{Argyres:2008rna}
P.~C. Argyres, A.~Awad, P.~Moomaw, and J.~Wittig, {\textquotedbl}{Holomorphic
  higher-derivative terms in supersymmetric effective actions},{\textquotedbl}
  {\em {Proceedings, 7th International Workshop on Supersymmetries and Quantum
  Symmetries (SQS'07): Dubna, Russia, July 30 - August 04, 2007}} (2008)
  267--274.
\url{https://inspirehep.net/record/1391563/files/P267.pdf}.

\bibitem{Seiberg:1994rs}
N.~Seiberg and E.~Witten, {\textquotedbl}{Electric - magnetic duality, monopole
  condensation, and confinement in $N=2$ supersymmetric Yang-Mills
  theory},{\textquotedbl}
  \href{http://dx.doi.org/10.1016/0550-3213(94)90124-4}{{\em Nucl. Phys.}
  {\bfseries B426} (1994) 19--52},
  \href{http://arxiv.org/abs/hep-th/9407087}{{\ttfamily arXiv:hep-th/9407087
  [hep-th]}}.
[Erratum: Nucl. Phys. B430, 485 (1994)].

\bibitem{Seiberg:1994aj}
N.~Seiberg and E.~Witten, {\textquotedbl}{Monopoles, duality and chiral
  symmetry breaking in $N=2$ supersymmetric QCD},{\textquotedbl}
  \href{http://dx.doi.org/10.1016/0550-3213(94)90214-3}{{\em Nucl. Phys.}
  {\bfseries B431} (1994) 484--550},
\href{http://arxiv.org/abs/hep-th/9408099}{{\ttfamily arXiv:hep-th/9408099
  [hep-th]}}.

\bibitem{Pestun:2007rz}
V.~Pestun, {\textquotedbl}{Localization of gauge theory on a four-sphere and
  supersymmetric Wilson loops},{\textquotedbl}
  \href{http://dx.doi.org/10.1007/s00220-012-1485-0}{{\em Commun. Math. Phys.}
  {\bfseries 313} (2012) 71--129},
\href{http://arxiv.org/abs/0712.2824}{{\ttfamily arXiv:0712.2824 [hep-th]}}.

\bibitem{Argyres:1995jj}
P.~C. Argyres and M.~R. Douglas, {\textquotedbl}{New phenomena in $SU(3)$
  supersymmetric gauge theory},{\textquotedbl}
  \href{http://dx.doi.org/10.1016/0550-3213(95)00281-V}{{\em Nucl. Phys.}
  {\bfseries B448} (1995) 93--126},
\href{http://arxiv.org/abs/hep-th/9505062}{{\ttfamily arXiv:hep-th/9505062
  [hep-th]}}.

\bibitem{Argyres:1995xn}
P.~C. Argyres, M.~R. Plesser, N.~Seiberg, and E.~Witten, {\textquotedbl}{New
  ${\mathcal{N}}=2$ superconformal field theories in
  four-dimensions},{\textquotedbl}
  \href{http://dx.doi.org/10.1016/0550-3213(95)00671-0}{{\em Nucl. Phys.}
  {\bfseries B461} (1996) 71--84},
\href{http://arxiv.org/abs/hep-th/9511154}{{\ttfamily arXiv:hep-th/9511154
  [hep-th]}}.

\bibitem{Xie:2012hs}
D.~Xie, {\textquotedbl}{General Argyres-Douglas Theory},{\textquotedbl}
  \href{http://dx.doi.org/10.1007/JHEP01(2013)100}{{\em JHEP} {\bfseries 01}
  (2013) 100},
\href{http://arxiv.org/abs/1204.2270}{{\ttfamily arXiv:1204.2270 [hep-th]}}.

\bibitem{Bobev:2013vta}
N.~Bobev, H.~Elvang, and T.~M. Olson, {\textquotedbl}{Dilaton effective action
  with $\mathcal{N} = 1$ supersymmetry},{\textquotedbl}
  \href{http://dx.doi.org/10.1007/JHEP04(2014)157}{{\em JHEP} {\bfseries 04}
  (2014) 157},
\href{http://arxiv.org/abs/1312.2925}{{\ttfamily arXiv:1312.2925 [hep-th]}}.

\bibitem{Aharony:2009gg}
O.~Aharony and E.~Karzbrun, {\textquotedbl}{On the effective action of
  confining strings},{\textquotedbl}
  \href{http://dx.doi.org/10.1088/1126-6708/2009/06/012}{{\em JHEP} {\bfseries
  06} (2009) 012},
\href{http://arxiv.org/abs/0903.1927}{{\ttfamily arXiv:0903.1927 [hep-th]}}.

\bibitem{Aharony:2011ga}
O.~Aharony, M.~Field, and N.~Klinghoffer, {\textquotedbl}{The effective string
  spectrum in the orthogonal gauge},{\textquotedbl}
  \href{http://dx.doi.org/10.1007/JHEP04(2012)048}{{\em JHEP} {\bfseries 04}
  (2012) 048},
\href{http://arxiv.org/abs/1111.5757}{{\ttfamily arXiv:1111.5757 [hep-th]}}.

\bibitem{Aharony:2013ipa}
O.~Aharony and Z.~Komargodski, {\textquotedbl}{The Effective Theory of Long
  Strings},{\textquotedbl}
  \href{http://dx.doi.org/10.1007/JHEP05(2013)118}{{\em JHEP} {\bfseries 05}
  (2013) 118},
\href{http://arxiv.org/abs/1302.6257}{{\ttfamily arXiv:1302.6257 [hep-th]}}.

\bibitem{Coleman:1969sm}
S.~R. Coleman, J.~Wess, and B.~Zumino, {\textquotedbl}{Structure of
  phenomenological Lagrangians. 1.},{\textquotedbl}
\href{http://dx.doi.org/10.1103/PhysRev.177.2239}{{\em Phys. Rev.} {\bfseries
  177} (1969) 2239--2247}.

\bibitem{Callan:1969sn}
C.~G. Callan, Jr., S.~R. Coleman, J.~Wess, and B.~Zumino,
  {\textquotedbl}{Structure of phenomenological Lagrangians.
  2.},{\textquotedbl}
\href{http://dx.doi.org/10.1103/PhysRev.177.2247}{{\em Phys. Rev.} {\bfseries
  177} (1969) 2247--2250}.

\bibitem{Kuzenko:2015jda}
S.~M. Kuzenko, J.~Novak, and G.~Tartaglino-Mazzucchelli, {\textquotedbl}{Higher
  derivative couplings and massive supergravity in three
  dimensions},{\textquotedbl}
  \href{http://dx.doi.org/10.1007/JHEP09(2015)081}{{\em JHEP} {\bfseries 09}
  (2015) 081},
\href{http://arxiv.org/abs/1506.09063}{{\ttfamily arXiv:1506.09063 [hep-th]}}.

\bibitem{Komargodski:2011vj}
Z.~Komargodski and A.~Schwimmer, {\textquotedbl}{On Renormalization Group Flows
  in Four Dimensions},{\textquotedbl}
  \href{http://dx.doi.org/10.1007/JHEP12(2011)099}{{\em JHEP} {\bfseries 12}
  (2011) 099},
\href{http://arxiv.org/abs/1107.3987}{{\ttfamily arXiv:1107.3987 [hep-th]}}.

\bibitem{Schwimmer:2010za}
A.~Schwimmer and S.~Theisen, {\textquotedbl}{Spontaneous Breaking of Conformal
  Invariance and Trace Anomaly Matching},{\textquotedbl}
  \href{http://dx.doi.org/10.1016/j.nuclphysb.2011.02.003}{{\em Nucl. Phys.}
  {\bfseries B847} (2011) 590--611},
\href{http://arxiv.org/abs/1011.0696}{{\ttfamily arXiv:1011.0696 [hep-th]}}.

\bibitem{Dine:1997nq}
M.~Dine and N.~Seiberg, {\textquotedbl}{Comments on higher derivative operators
  in some SUSY field theories},{\textquotedbl}
  \href{http://dx.doi.org/10.1016/S0370-2693(97)00899-X}{{\em Phys. Lett.}
  {\bfseries B409} (1997) 239--244},
\href{http://arxiv.org/abs/hep-th/9705057}{{\ttfamily arXiv:hep-th/9705057
  [hep-th]}}.

\bibitem{Luty:2012ww}
M.~A. Luty, J.~Polchinski, and R.~Rattazzi, {\textquotedbl}{The $a$-theorem and
  the Asymptotics of 4D Quantum Field Theory},{\textquotedbl}
  \href{http://dx.doi.org/10.1007/JHEP01(2013)152}{{\em JHEP} {\bfseries 01}
  (2013) 152},
\href{http://arxiv.org/abs/1204.5221}{{\ttfamily arXiv:1204.5221 [hep-th]}}.

\bibitem{Nekrasov:2002qd}
N.~A. Nekrasov, {\textquotedbl}{Seiberg-Witten prepotential from instanton
  counting},{\textquotedbl}
  \href{http://dx.doi.org/10.4310/ATMP.2003.v7.n5.a4}{{\em Adv. Theor. Math.
  Phys.} {\bfseries 7} no.~5, (2003) 831--864},
\href{http://arxiv.org/abs/hep-th/0206161}{{\ttfamily arXiv:hep-th/0206161
  [hep-th]}}.

\bibitem{Alday:2009aq}
L.~F. Alday, D.~Gaiotto, and Y.~Tachikawa, {\textquotedbl}{Liouville
  Correlation Functions from Four-dimensional Gauge Theories},{\textquotedbl}
  \href{http://dx.doi.org/10.1007/s11005-010-0369-5}{{\em Lett. Math. Phys.}
  {\bfseries 91} (2010) 167--197},
\href{http://arxiv.org/abs/0906.3219}{{\ttfamily arXiv:0906.3219 [hep-th]}}.

\bibitem{barnes}
E.~W. Barnes, {\textquotedbl}{The Theory of the Double Gamma
  Function},{\textquotedbl}
  \href{http://dx.doi.org/10.1098/rsta.1901.0006}{{\em Phil. Trans. Roy. Soc.
  Lond. A} {\bfseries 196} (1901) 265--387}.

\bibitem{Aniceto:2014hoa}
I.~Aniceto, J.~G. Russo, and R.~Schiappa, {\textquotedbl}{Resurgent Analysis of
  Localizable Observables in Supersymmetric Gauge Theories},{\textquotedbl}
  \href{http://dx.doi.org/10.1007/JHEP03(2015)172}{{\em JHEP} {\bfseries 03}
  (2015) 172},
\href{http://arxiv.org/abs/1410.5834}{{\ttfamily arXiv:1410.5834 [hep-th]}}.

\bibitem{Shapere:2008zf}
A.~D. Shapere and Y.~Tachikawa, {\textquotedbl}{Central charges of
  ${\mathcal{N}}=2$ superconformal field theories in four
  dimensions},{\textquotedbl}
  \href{http://dx.doi.org/10.1088/1126-6708/2008/09/109}{{\em JHEP} {\bfseries
  09} (2008) 109},
\href{http://arxiv.org/abs/0804.1957}{{\ttfamily arXiv:0804.1957 [hep-th]}}.

\end{thebibliography}\endgroup

\end{document}